\documentclass[preprint,aps,nopacs,preprintnumbers,amsmath,amssymb]{revtex4}

\usepackage{amsmath}

\usepackage{graphicx}
\usepackage{dcolumn}
\usepackage{bm}
\usepackage{color}

\usepackage{amsmath}
\usepackage{amssymb}

\def\be{\begin{equation}}
\def\en{\end{equation}}
\def\bea{\begin{eqnarray}}
\def\ena{\end{eqnarray}}
\def\p{\partial}
\def\ep{\epsilon}
\def\gs{\gtrsim}
\def\ls{\lesssim}

\newcommand{\av}[1]{\langle{#1}\rangle}

\newcommand{\bi}[1]{\mbox{\boldmath$#1$}}

\newcommand{\ROcite}[1]{Ref.\onlinecite{#1}}
\newcommand{\ov}[1]{\overline{#1}}

\begin{document}
\title{Theory of central peak and acoustic anomaly 
in cubic BaTiO$_3$\\ close to   ferroelectric transition 
}  
\author{Akira Onuki\footnote{e-mail: onuki@scphys.kyoto-u.ac.jp 
(corresponding author)}}

\affiliation{
 Department of Physics, Kyoto University, Kyoto 606-8502, Japan 
}


\date{\today}

\begin{abstract} 
We present  a Ginzburg-Landau  theory on  statics and dynamics   
of BaTiO$_3$-type ferroelectrics in 
 the paraelectric phase 
with the cubic structure, where the order parameter 
is the polarization $\bi p$. 
Unique effects are caused by the  electrostrictive (ES) coupling between 
 ${\bi p}$ and  the elastic displacement $\bi u$.    
We show that the ES coupling  gives rise to 
a central peak in the Fourier-Laplace transform 
of the displacement  time-correlation function 
at small wave numbers. 
It emerges and grows with a narrow width 
as the transition is approached. 
Such central peaks  have long been   observed  
in a number of    scattering 
experiments in various ferroelectrics, 
but their origin has not been well 
understood. From the acoustic part of the displacement dynamic 
correlation  we obtain  the frequency-dependent 
elastic moduli $C_{11}^*(\omega)$, $C_{12}^*(\omega)$, 
and $C_{44}^*(\omega)$, whose singular parts 
arise from the ES coupling, 
We then calculate 
the singular sound velocity  and 
 attenuation.  In the central peak and the 
elastic moduli,  the frequency $\omega$ appears 
in the  scaled form  $\omega\tau_D$,  where  $\tau_D$ 
is the Debye relaxation  time in the frequency-dependent 
dielectric constant.\\  
\vspace{4mm}
 \end{abstract}
\maketitle

{\bf 1. Introduction}\\ 
Ferroelectric transitions  have  been studied     
 extensively in theories and experiments 
\cite{Wen,Mo,Landau-e,Mitsui,Dev,Shirane,Str,Tagan}.  
where the static  dielectric constant $\ep$ grows 
as $(T-T_0)^{-1}$ obeying the Curie-Weiss law in 
the paraelectric phase 
as a function of the temperature $T$. 
Many such systems including barium titanade 
BaTiO$_3$ undergo weakly first-order transitions 
with structure changes \cite{Samara,Clarke,Decker,Mazur},   
where  the transition temperature  $T_c$ is  slightly above  
the psudocritical temperature $T_0$. 
 In the ordered phase with  $T<T_c$, 
   spontaneous polarization  appears resulting in domain formation. 
  To understand the   transition behavior, use has been made of 
the Landau theory \cite{Binder,Gin,Litt,Str,Onukib,Cross1,Chen2,Nam}, 
where the free energy of the polarization density $\bi p$ 
  is a polynomial of  $\bi p$  with the gradient term 
and the electric energy added.  n
As a unique feature,  the electric field $\bi E$ 
 consists of the contribution 
from  the electrode charges and that from 
the interior dipoles, where the latter largely cancels 
the former and gives rise to the dipole-dipole 
interaction. In  particular, Kretschmer and  Binder\cite{Binder} 
accounted for  this  depolarization effect  
in their Landau free energy  
to  calculate  the polarization profile in  film samples. 

The ferroelectric transitions   are strongly influenced 
by the solid elasticity, where  $\bi p$ 
and  the elastic strains $e_{\alpha\beta}$ 
($\alpha,\beta=x,y,z)$ are coupled in the free energy 
 \cite{Dev,Tagan,Shirane}. 
The coupling  is of  the bilinear form 
${\sum}_{\alpha,\beta,\gamma} 
d_{\alpha\beta\gamma}p_\alpha e_{\beta\gamma}$  
in  piezoelectric systems, where  $d_{\alpha\beta\gamma}$ are  constants.  
A famous example is  uniaxial KH$_2$PO$_4$ (KDP), where   
the elastic modulus $C_{66}$ tends to $0$  
and becomes dependent on the  frequency $\omega$ 
 on approaching the transition in the 
paraelectric phase \cite{Cum1,Cum,Gar,Lito}. 
Such strong elastic  anomaly  can be  explained 
 by the Landau-Khalatnikov theory \cite{Khala} 
with    the long-range dipolar interaction 
included \cite{Kri2,Kri1}.   
  On  the other hand, 
in systems with a centrosymmetric crystal structure,   
the  coupling  is  electrostrictive (ES)  
in the paraelectric phase, where 
it  is of the third order form  
${\sum}_{\alpha,\beta,\gamma,\nu} 
d_{\alpha\beta\gamma\nu}p_{\alpha}p_\beta 
 e_{\gamma\nu}$ in the lowest order with $d_{\alpha\beta\gamma\gamma}$  
being constants \cite{Bell,TYama,Uwe,Hinka,Uchino,Cross,Chen1,Renoud}.
This coupling becomes piezoelectric with appearance of nonvanishing 
average  polarization. 
Famous examples are  BaTiO$_3$ 
and   uniaxial triglycine sulfate (TGS) \cite{Mitsui}, 
whose crystal structures are  cubic and 
monoclinic, respectively,  in paraelectric states. 
Theories on the near-critical sound absorption 
were presented for BaTiO$_3$ \cite{Leva2,Dvorak} 
 and TGS \cite{Leva1,Mina} many years ago.
The acoustic anomaly in  electrostrictive  systems 
\cite{Lito1,Kashida,Ko1} 
are much smaller in magnitudes than that in piezoelectic systems.

In various ferroelectrics, 
   central peals have been observed  in the dynamical 
scattering amplitudes  
\cite{Shira0,Shira1,Sha,Top,Riste,Ly,Ly1,Ko,Ko3,To,Cum2,Cum1,Cow,Ge,Y1,Y,Y2}, 
which  emerge and grow with a narrow width as $T\to T_c$ 
 both  in piezoelectric   and electrostrictive systems. 
In    strontium  titanade SrTIO$_3$,  however, 
the central peak was observed  
 for the  zone-boundary $(1/2,1/2,3/2)$   wave vector 
near  the $105~$K  antiferrodistortive,  second-order phase 
transition,  where the order parameter is the staggered rotation 
angles of the oxygen octahedra 
but the polarization fluctuations are also enhanced  
 with $\ep\sim 10^3$.
  Previously, the  origin of the central peaks 
was ascribed to the thermal entropy 
diffusion (Rayleigh scattering) \cite{Y2,Cum1,Ly1}, 
to  impurities \cite{Varma}, 
and to droplets or small 
domains \cite{Cook}. In  light scattering,  
the  dielectric tensor at optical frequencies 
  depends on the polarization and/or 
 the strains \cite{Cum1,Ginzburg,To,Burns,Vacher}, 
which can  result  in the central peaks.

From our viewpoint, the central peaks  are 
  produced by slowly relaxing components in the stress tensor, which 
can arise  from the piezoelectric or electrostrictive (ES) coupling 
without impurities, vacancies, and small domains. 
We gain this picture from our calculation 
of the time-correlation of  $\bi u$ at  small wave vector $\bi q$ 
 with the ES coupling for BaTiO$_3$. 
We shall see that its temporal  Fourier transform 
has  a narrow peak centered at $\omega=0$, which 
exhibits   the characteristic feature of 
the  central peak of BaTiO$_3$ \cite{Shira1,Ko,Ko3}. 
Thus, it should  constitue  an essential  
part of   the   central peak in the neutron and light 
scattering amplitudes \cite{Coch}. We note that the 
longitudinal acoustic modes exhibit critical anomaly 
in BaTiO$_3$. In this calculation  we also 
obtain the frequency-dependent elastic moduli $C_{ij}^*(\omega)$, 
where  $C_{11}^*(\omega)$ 
and $C_{12}^*(\omega)$ exhibit significant critical anomaly 
but $C_{44}^*(\omega)$ is nearly 
nonsingular \cite{Ko1}.

The organization of this paper is as follows. 
In Sec.II, the statics in the Landau theory will be discussed.  
In Sec.III, the equal-time polarization correlations will  be calculated.    
 In Sec.IV, we  will  calculate  the time-correlations of the 
polarization and displacement fluctuations, 
where the frequency-dependent elastic moduli   
and the central peak will be derived.

\noindent{\bf 2. Statics}\\
\vspace{2mm}
\noindent
{\bf 2.1.  Free energy of $\bi p$ and $\bi u$}\\ 
 We consider a  BaTiO$_3$-type  cubic  system   
slightly above the  transition 
 without free charges. The polarization $\bi p$ exhibits 
critical fluctuations.  The sample   is confined  
in a  $L\times L\times H$ cell 
with metal electrodes  at $ z=0$ and $H$. 
Here, $d \ll  H\ll L$,  where $d$ is the lattice constant. 
The sample volume is $V=L^2H$.  We write the   Maxwell electric field as 
 ${\bi E}({\bi r})=-\nabla\Phi({\bi r})$  with ${\bi E}={\bi 0}$  
in the electrodes, where 
the  potential  $\Phi$   arises 
 from the electrode charges and the interior dipoles.
The electric induction is  
${\bi D}=  {\bi E}+ 4\pi {\bi p}$ with  
 $\nabla\cdot{\bi D}=0$ in the sample. 
In this geometry, the  net  surface charges at $z=0$ and $z=H$ 
are written as $Q_0$ and $-Q_0$, respectively, 
while the potential  at $z=0$ and $z=H$ 
can be   set equal to $\Phi_a$ and $0$, respectively 
\cite{OnukiD,Van,Sprik1,Sprik,Onuki2025}. 

The  free energy functional for $\bi p$ and the 
elastic displacement $\bi u$  is given by 
\be 
{\cal F}(\{{\bi p}\},\{{\bi u}\})=\int d{\bi r}\Big[\frac{1}{8\pi}|{\bi E}|^2  
 +f_{\rm B}+ f_{\rm el}+f_{\rm int}+f_T\Big],
\en 
where $\int d{\bi r}(\cdots)$ 
is the sample integration and   $\int d{\bi r} |{\bi E}|^2 /8\pi$ 
is the  electric energy \cite{Sprik1,Sprik,Onuki2025,OnukiD,Landau-e,Van}.
To see how the electric energy is related to $\bi p$, 
 let us   superimpose   small changes,  
 $\delta {\bi p}$,  $\delta {\bi E}$, and $\delta Q_0$  
 on  $\bi p$,  $\bi E$, and $Q_0$, respectively. The   
 incremental  change in the electric energy becomes     
\be 
 \int d{\bi r}{\bi E}\cdot\delta{\bi E}/4\pi=
\Phi_a\delta Q_0 -  \int d{\bi r}{\bi E}\cdot\delta{\bi p}, 
\en 
where  the first term vanishes 
at  fixed $Q_0$ or at $\Phi_a=0$. Thus, we assume 
either of these two conditions. 
Then, the functional derivative of the electric energy 
with respect to $\bi p$ is simply given by $-{\bi E}$.  
In this paper, we   
 omitt   the surface free energy in Eq.(1) \cite{Binder} 
and the effect of the surface layers \cite{Onuki2025}  
will be discussed briefly in 
the last paragraphs in Secs.2.4 and 4.1.  

The appropriate free energy 
at   a fixed  $\Phi_a$ 
is given by the Legendre transform    
${\tilde {\cal F}}= {\cal F}-\Phi_{a}  Q_0=
{\cal F}-\int d{\bi r}{\bi E}
\cdot {\bi D}/4\pi$  \cite{Landau-e,Onuki2025,OnukiD},
which leads to 
 $(\delta {\cal {\tilde F}}/\delta{\bi p})_{\Phi_a.{\bi u}}=
(\delta {\cal F}/\delta{\bi p})_{Q_0,{\bi u}}$ (see Eq.(19)).  
In $\tilde{\cal{F}}$,  the electric energy density is given by  
 $-|{\bi E}|^2/8\pi-{\bi p}\cdot{\bi E}$, which 
gives rise to  
the   one-dimensional free energy 
in the Kretchmer-Binder theory (see their Eq.(12)) \cite{Binder}. 

\vspace{2mm}
\noindent
{\bf 2.2.  Bulk free energy of $\bi p$ and first-order 
ferroelectric transition}\\ 
In Eq.(1),  $f_{\rm B}$ is the bulk free energy density of $\bi p$, 
where one of the easy axes   is along the $z$ axis. 
Up to the sixth order terms it is written 
as \cite{Bell,Dev,Shirane,Nam,Cross,Chen1,Chen2,Cross1} 
\bea  
&&\hspace{-19mm}f_{\rm B}= 
 \frac{1}{2} A |{\bi p}|^2+
\alpha_{11}^0 {\sum}_\alpha p_\alpha^4+ 
\alpha_{12}^0 (p_x^2 p_y^2  +p_y^2 p_z^2+ p_z^2 p_x^2)\nonumber \\
&&\hspace{-9mm}
+ {\sum}_{\alpha,\beta,\gamma}\alpha^0_{\alpha\beta\gamma} 
p_\alpha^2 p_\beta^2p_\gamma^2+\frac{1}{2} C |{\nabla{\bi p}}|^2,
\ena 
where   $\alpha^0_{11}$, $\alpha^0_{12}$, and $a^0_{\alpha\beta\gamma}$   
are constants.   Hereafter, the Greek indices refer to the Cartesian coordinates, $x$, $y$, and $z$.  The coefficient  $A$ is related to the 
temperature $T$  as  
\be 
A= a_1(T/T_0-1). 
\en  
For a  weakly first-order transition, 
  $T_0$  is  the psudocritical temperaure 
slightly  below the   first-order transition temperature  $T_c$.
For BaTiO$_3$, the  fourth order terms  are  small 
 and    the sixth order terms   are indispensable 
 in  describing  the phase behavior. See Appendix A, 
where the fourth-order terms produced by the ES interaction 
will be calculated. 

As  Eqs.(2) and (3) indicate, 
 the   dielectric susceptibility $\chi$ and 
the dielectric  constant $\ep=1+4\pi\chi$ 
 in the linear regime are related to $A$ as   
 \be 
\chi=1/A, ~~~
\ep=1+4\pi/A \cong  (4\pi /a_1)/(T/T_0-1), 
\en  
which is  the Curie-Weiss law  for $0<A\ll 1$.
 The surface effect on the dielectric 
constant will  be discussed in the last paragraph 
in Sec.2.4. 
  
At low ambient pressures,   BaTiO$_3$ undergoes   a 
weakly first-order ferroelectric  transition 
with a cubic-to-tetragonal structure change 
at $T_c \cong 400$K. The transition 
becomes second-order with increasing the pressure 
$(\gs 35~{\rm Kbar}= 3.5{\rm GPa})$ \cite{Decker,Samara,Clarke}. 
This can be explained by the pressure-dependence 
of   the coefficients in the Landau model. 
In particular, the experimental values of $a_1$, $T_0$, and $T_c$ 
at low pressures depend on the experimental groups. 
We mention two examples:  
\setcounter{equation}{0}
\bea 
&& \hspace{18mm} a_1=  1.8\times 10^{-2},~~ T_0= 384~{\rm K},
~~T_c-T_0= 17~{\rm K}, \nonumber \hspace{4.9cm}(6a) 
\\
&&\hspace{18mm}  a_1=3.6\times 10^{-2}, ~~T_0=388~{\rm K},~~~
   T_c-T_0= 10~{\rm K},  \nonumber \hspace{4.8cm}(6b) 
\ena  
where we use the data 
by  Kashida {\it et al.} \cite{Kashida} in Eq.(6a) 
and those by  Li {\it et al.}  \cite{Cross} in Eq.(6b).  
\setcounter{equation}{6}
Here,  $a_1\ll 1$, whose implication was 
discussed by Ginzburg \cite{Gin}. 
The $a_1$ in Eq.(6b) is  twice 
larger  than that in Eq.(6a), but   both  
give  $\ep\sim 1.5\times 10^4$ and 
$A\sim 10^{-3}$ at $T=T_c$,  
which are the maximum of $\ep$ and the minimum of $A$, respectively,  
in equilibrium paraelectric  BaTiO$_3$. 
It is worth noting that $a_1=1.24$ for  
TGS \cite{Mitsui}  and $a_1=0.41$ for   
KDP \cite{Gar}.

 In the gradient free energy we set   
 $|\nabla{\bi p}|^2= {\sum}_{\alpha,\beta}(\nabla_\alpha p_\beta)^2$,  
where  $\nabla_\alpha$ is   the $\alpha$-th component of $\nabla=
\p/\p{\bi r}$. This form of  the gradient free energy 
much simplifies the calculations to follow, 
though it can be anisotropic generally 
(see Appendix B) \cite{Nam,Cross,Chen1,Chen2,Cross1}. 
 From Eq.(3) we  find the growing mean-field 
 correlation length given by  
\bea 
&& \xi= (C/A)^{1/2}=
\xi_0 (T/T_c-1)^{-1/2},~~\xi_0= (C/a_1)^{1/2} .
\ena 
where $\xi_0\sim  10 \sqrt{C}$ from Eq.(6). 
In  Eq.(50) below,  
we will have   $\sqrt{C}\cong 1.5~{\rm \AA}$
      for  BaTiO$_3$. Thus, we  find  
 $\xi_0 \sim 10~{\rm \AA}$, 
which is  longer than 
 the lattice constant $d=4~{\rm \AA}$.  

\vspace{2mm}
\noindent
{\bf 2.3.  Elastic,  electrostrictive, and thermoelastic  
free energies }\\ 
  The $f_{\rm el}$ in Eq.(1) is the elastic free energy 
of cubic solids written   as 
\bea
&&\hspace{-10mm}f_{\rm el}=\frac{1}{2}{\sum}_{\alpha,\beta} 
\sigma_{\alpha \beta}\epsilon_{\alpha \beta} 
= \frac{K}{2}(\nabla\cdot{\bi u}|^2 + \frac{c_{11}-c_{12}}{2}
{\sum}_\alpha \Big(e_{\alpha\alpha}- \frac{1}{3}{\nabla\cdot{\bi u}}\Big)^2
+ c_{44}{\sum}_{\alpha\neq \beta}e_{\alpha\beta}^2
\ena 
where    $\epsilon_{\alpha\beta}=(\nabla_\alpha u_\beta+\nabla_\beta u_\alpha)/2$ are the strains  and 
 $\sigma_{\alpha\beta}$ are the elastic 
stress components, 
\bea 
&&\hspace{-10mm} \sigma_{\alpha\beta}=
\Big[(c_{11}-c_{12})\epsilon_{\alpha\alpha} 
+ c_{12}{\sum}_\gamma \epsilon_{\gamma\gamma}\Big]\delta_{\alpha\beta}
+2c_{44}(\epsilon_{\alpha\beta}- 
\epsilon_{\alpha \alpha}\delta_{\alpha\beta}). 
\ena 
The  $c_{ij}$ are  the  bare (unrenormalized) 
 elastic moduli and  $K=(c_{11}+2c_{12})/3$ 
is   the bare bulk modulus, We suppose a  
reference equilibrium state 
above $T_c$ in the stress-free boundary 
condition (or at a fixed low ambient pressure) without applied field,  
where the thermal averages of the stains $\av{\ep_{\alpha\beta}}$ 
 vanish. 
At long wavelengths,    the mass density deviation  $\delta\rho$ is 
  related  to the dilation strain  
$\nabla\cdot{\bi u}$ as  \cite{Landau-ela} 
\be 
\delta\rho = - \rho \nabla\cdot{\bi u}=-\rho{\sum}_\gamma 
\epsilon_{\gamma\gamma},
\en 
where $\rho$ is  the average mass density 
and no  vacancies are assumed in the crystal.

\narrowtext
\begin{table}
\caption{Coefficients for cubic  BaTiO$_3$.
First line: $c_{ij}$ (in GPa)\cite{Chen1} and    
$Q_{11}, Q_{12},Q_{44}$,  $Q_{11}-Q_{12}$, 
 and $Q_{11}+2Q_{12}$  (in m$^4/$C$^2$)\cite{Chen1}, 
where $1~{\rm m}^4/{\rm C}^2\cong 0.11/$GPa.
Second line: $g_{ij}$, and $M_{11}, M_{12}, M_{44}$,  $
M_{11}-M_{12}$, and $M_{11}+2M_{12}$, 
where $M_{ij}$ will be related to $g_{ij}$  in Eq.(41).}
\label{tab:2d}
\begin{tabular}{cccccccc}
\hline 
$c_{11}~~$&$c_{12}~~$&$c_{44}~~$&$Q_{11}~~$&
$Q_{12}~~$&$Q_{44}~~$&$Q_{11}-Q_{12}~~$&$Q_{11}+2Q_{12}$\\
\hline
305~~  &106~~ & 128~~ & 0.115~~ &-0.033~~ &0.041~~&0.148~~&0.049 \\
\hline
\hline 
$g_{11}~~$&$g_{12}~~$&$g_{44}~~$&$M_{11}~~$&
$M_{12}~~$&$M_{44}~~$&$M_{11}-M_{12}~~$&$M_{11}+2M_{12}$\\
\hline
4.28~~  &-0.71~~ & 0.59~~ & 4.77~~ & -1.04~~ & 0.041~~&5.81&2.69 \\
\hline
\end{tabular}
\end{table}

The  $f_{\rm int}$ in Eq.(1) represents 
  the ES  coupling as   
\bea 
&&\hspace{-11mm} f_{\rm int}= 
- {\sum}_{\alpha,\beta} {\Pi}_{\alpha\beta}^{\rm s} \epsilon_{\alpha\beta}
=- {\sum}_{\alpha,\beta} {e}_{\alpha\beta}^{\rm s} \sigma_{\alpha\beta}. 
\ena  
Here, $\{ {\Pi}_{\alpha\beta}^{\rm s}\}$ is the singular 
stress tensor and $\{\epsilon_{\alpha\beta}^{\rm s}\}$ 
 is  the singular  strain tensor, which are  
 bilinear in $\bi p$. For cubic solids 
they are expressed  as  \cite{Cross,Bell,Uwe,TYama,Hinka,Chen1,Nam} 
\bea 
&&\hspace{-9mm}\Pi_{\alpha\beta}^{\rm s}=  [(g_{11}-g_{12}) p_\alpha^2 
+ g_{12}|{\bi p}|^2]\delta_{\alpha\beta}+g_{44} (p_\alpha p_\beta-
p_\alpha^2 \delta_{\alpha \beta}) ,
\\
 &&\hspace{-8mm} 
e^{\rm s}_{\alpha\beta}\hspace{-0.5mm}
=\hspace{-0.5mm}  [(Q_{11}-Q_{12}) p_\alpha^2 
+ \hspace{-0.5mm}
Q_{12} |{\bi p}|^2]\delta_{\alpha\beta}\hspace{-0.7mm}
+\frac{1}{2}
Q_{44} (p_\alpha p_\beta-p_\alpha^2 \delta_{\alpha \beta}) ,    
\ena 
where $g_{ij} $ and $Q_{ij}$ are constants. 
In Voigt notation we have $\ep_1^{\rm s}= \ep^{\rm s}_{xx}$, 
$\ep_2^{\rm s}= \ep^{\rm s}_{yy}$, 
$\ep_3^{\rm s}= \ep^{\rm s}_{zz}$, 
$\ep_4^{\rm s}=2e^{\rm s}_{yz}$, 
$\ep_5^{\rm s}=2e^{\rm s}_{zx}$, and 
 $\ep_6^{\rm s}=2 e^{\rm s}_{xy}$ \cite{Chen1,Nam}.  
Since   Eq.(11)  holds for any  $\epsilon_{\alpha\beta}$, 
we find    
\be
Q_{11}- Q_{12}= \frac{g_{11}- g_{12}}{c_{11}-c_{12}},~~~ 
Q_{11}+2 Q_{12}= \frac{g_{11}+2 g_{12}}{c_{11}+2c_{12}},~~~ 
Q_{44}=\frac{g_{44}}{c_{44}}, ~~
\en 
To derive these relations 
 we set   $\Pi_{\alpha\alpha}^{\rm s}= 
(g_{11}-g_{12}) {\cal D}_\alpha 
+ (g_{11}+2g_{12})|{\bi p}|^2/3$ and 
$e^{\rm s}_{\alpha\alpha}=(Q_{11}-Q_{12}) {\cal D}_\alpha 
+ (Q_{11}+2Q_{12})|{\bi p}|^2/3$ 
 with ${\cal D}_\alpha =
 p_\alpha^2 - |{\bi p}|^2 /3$ (see Eqs.(C4)-(C7)).

In Table 1, we present    $c_{ij}$,  $Q_{ij}$, 
and $g_{ij}$ using theoretical  values 
by Chen's group\cite{Chen1}  
for cubic BaTiO$_3$, 
where $M_{ij}$ will be  introduced in Sec.III. 
Their   $Q_{ij}$ values  are 
consistent with  measured values\cite{Bell,Hinka,TYama}. 
The  $Q_{ij}$ were also measured  for KaTiO$_3$ \cite{Uwe}. 
They have been obtained from 
shape changes of single crystals 
 with uniform  polarization in unclamped samples. 
See Appendix A for more discussions on this 
aspect. 

In addition, 
Landau and Lifshitz introduced 
 the thermoelastic free energy  density 
 \cite{Landau-ela}, 
\be 
f_T= -K\alpha_P (T-T_{\rm ex})\nabla\cdot{\bi u},  
\en	 
where $T_{\rm ex}$ is the  reference  temperature above $T_c$, 
$K= (c_{11}+2c_{12})/3$  is the bulk modulus,  
and  $\alpha_P =-(\p \rho/\p T)_P/\rho$  is the 
thermal expansion coefficient. In $\alpha_P$ the ambient 
pressure $P$ is fixed, which is equal to $P_{\rm ex}$ 
in the reference state. This   $P$ should not be 
confused with the polarization $\bi p$. 
Then, a temperature change  $\delta T$ from $T_{\rm ex}$ 
induces a density change  
$\delta\rho=-\rho \alpha_P \delta T$ in equilibrium at fixed $P$. 
The total stress tensor is written as       
\be 
\Pi_{\alpha\beta}^{\rm tot}=
-\sigma_{\alpha\beta}+\Pi^{\rm s}_{\alpha\beta} 
+[K\alpha_P (T-T_{\rm ex})+P_{\rm ex}] \delta_{\alpha\beta}. 
\en 
We also define the internal pressure $\hat P$  by 
\be 
{\hat P}=  {\sum}_\alpha 
\Pi_{\alpha\alpha}^{\rm tot}/3=  -
 K\nabla\cdot{\bi u}  +(g_{11}+2g_{12})|{\bi p}|^2/3.
+K\alpha_P (T-T_{\rm ex})+P_{\rm ex}.
\en 
We have the thermal averages 
 $\av{\Pi_{\alpha\beta}^{\rm tot}}= \av{{\hat P}}
\delta_{\alpha\beta}=
P\delta_{\alpha\beta}$  in equilibrium unclamped samples.  
 
In the paraelectric phase  without  anisotropic strains  we  
 can define the thermodynamic entropy $s(T,P)$ per unit mass 
and    use the Maxwell relations \cite{Cohen,Gas,Gas1}. 
Then, we  find     
  the thermodynamic relations 
 $K= \rho(\p P/\p \rho)_T $ 
and $K\alpha_p= (\p P/\p T)_\rho$  \cite{Landau-ela}.  
For example, in longitudinal sounds, $T$   adiabatically varies  
by  $-\rho(\p T/\p \rho)_s  \nabla\cdot{\bi u}$. Then, in Eq.(17), 
the isothermal bulk modulus $K$   is replaced by  the adiabatic one 
 \cite{Landau-ela}, 
\be 
K_{\rm ad}=K[1 -(\p \rho/\p T)_P(\p T/\p \rho)_s]
=K\gamma= \rho (\p P/\p\rho)_s,
\en 
where    $c_{11}-c_{12}$ and $c_{44}$ are 
unchanged.  Here, $\gamma=C_P/C_V$ is 
the specific heat ratio, which is very close to 1 
for most solids (including BaTiO$_3$). Thus, we will set $K_{\rm ad}=K$ 
in Sec.IV.

\vspace{2mm}
\noindent
{\bf 2.4.  Functional derivatives  and  dielectric 
constant of films   }\\ 
From  Eqs.(1),  (2), and (16)  
 the functional derivatives of $\cal F$ 
are  given by   
\bea 
&&\hspace{-8mm} \Big(\frac{\delta{\cal F} }{\delta {\bi p}}\Big)_{{\bi u}}=
-{\bi E} -C\nabla^2{\bi  p}+ 
\Big(\frac{\p (f_{\rm B}+f_{\rm int})}{\p {\bi p}}\Big)_{\bi u} ,\\
&&
\hspace{-8mm}\Big(\frac{\delta {\cal F}}{\delta u_\alpha}\Big)_{{\bi p}}
= {\sum}_\beta \nabla_\beta \Pi^{\rm tot}_{\alpha\beta}
= -{\sum}_\beta \nabla_\beta \sigma_{\alpha\beta}
+{\sum}_\beta \nabla_\beta \Pi^{\rm s}_{\alpha\beta}
+ K\alpha_P \nabla_\alpha T, 
\ena  
which  vanish for the  equilibrium 
averages of ${\bi p}$ and ${\bi u}$. 
The  last term in Eq.(20) is negligible 
for homogeneous $T$ and 
in discussing  sounds (see below Eq.(18)).

We also  remark on  the finite-size  effect 
in the observed dielectric constant $\ep_{\rm eff} $ 
of films \cite{Van1,Sharma,Ab,Chin,Sekido,Sch,Mitsui,Gregg}. 
Let us suppose   thin surface layers   with  a thickness $d_s (\ll H)$ 
and a low dielectric constant $\ep_s (\ll \ep)$ 
in the presence of surface charges 
 $\pm  Q_0$ at  $z=0$ and $H$. In this situation,  
 the electric field   in the layers $E_s$  is 
much larger than  the bulk field $E_b$  from  
$\ep_s E_s= \ep E_b=D_z=  4\pi Q_0/L^2$. 
Defining $\ep_{\rm eff}$ by 
 $\Phi_a= \int_0^H dz E_z(z)=HD_z/\ep_{\rm eff}$, we obtain  
\be 
1/\ep_{\rm eff} = [(H-2d)/\ep+2d/\ep_s]/H= 1/\ep+ 2\ell_s/H,
\en 
 where  $\ell_s= d_s(/\ep_s-1/\ep)\cong d_s/\ep_s$, 
 $\ep_{\rm eff} \cong  H/2\ell_s$ 
for $\ep> H/2\ell_s$, and  $E_b= \Phi_a/(H + 2\ep\ell_s)$. 
See Eq.(60) for an effective frequency-dependent 
dielectric constant of films. 
 We note that Eq.(21) 
 can be derived from    the 
Kretschmer-Binder surface free energy in the Ginzburg-Landau 
scheme \cite{Binder,Tagan2}. 
For ferroelectrics, 
the observed size of $\ell_s$ depends on the  experimental 
conditions and its   origin has long been 
discussed\cite{Van1,Sharma,Gregg}. In electrochemistry 
the combination 
$C_s=1/4\pi\ell_s= \ep_s/4\pi d_s$ is  
called the surface capacitance  \cite{Hamann}.
 We point out that  the effective dielectric constant of 
 polar fluids depends on the film thickness $H$ 
 in the same manner\cite{Onuki2025,Takae,Laage,Cox1} 
due to  the Stern layers with a microscopic width 
 at metal-water interfaces \cite{Hamann}, which 
was confirmed experimentally 
for nanoconfined liquid water \cite{ZZ}.

\vspace{2mm}
\noindent{\bf 3. Correlations in  Fourier space}\\
\vspace{2mm} 
{\bf 3.1. Ginzburg criterion 
 and mean-field approximation}\\
 In  calculating  the polarization correlations 
for $T>T_c$, we can treat the fourth order terms in Eq.(1) 
as a perturbation. Their contributions are small corrections  for 
$A>A_{\rm Gi}$ (see   Eq.(4.125) in \ROcite{Onukib}). The lower bound 
$A_{\rm Gi}$ is calculated as \cite{Gin,Leva2, Onukib,Amit}
\be 
A_{\rm Gi}= ({a_{\rm 4th}k_BT}/{2\pi^2} )^2/C^3,  
\en  
where   $a_{\rm 4th}$ is the largest  value of the 
 coefficients of the fourth order terms in $f_{\rm B}$. 
For  $A>A_{\rm Gi}$ we can use the mean-field theory, 
where the polarization  fluctuations 
obey the Gaussian distribution. 
For BaTiO$_3$,  $A_{\rm Gi} $  is   extremely small \cite{Gin}. 
In fact,    $A_{\rm Gi}\sim  10^{-6}$ 
if   $a_{\rm 4th}$ is set equal to $ a_{12}$  
in the work by Li {\it et al.} \cite{Cross}. 
It increases up to $  10^{-5}$ 
if   $a_{\rm 4th}$ is equated to $ \beta_1$ in Eq.(A5) 
in Appendix A. Here, we assume  
$\sqrt{C}\cong 1.5~{\rm \AA}$ (see  Eq.(50)). 
It is known  that   BaTi0$_3$ at low pressures 
 is  close to a  tricritical point  
in the $P$-$T$ phase diagram 
  \cite{Mazur,Decker,Samara,Clarke}. 

\vspace{2mm} 
\noindent{\bf 3.2. Fourier components and polarization correlation}\\
We consider   the thermal  fluctuations of $\bi p$ and $\bi u$ 
 in the bulk region far from the surfaces 
without applied field, 
where the translational symmetry nearly holds. 
Their  statistical distribution is 
proportional to $\exp(-{\cal F}(\{{\bi p}\},\{{\bi u}\})/k_BT)$. 
Hereafter,  $\av{\cdots}$ 
denotes the equilibrium average.
 We calculate the correlations for 
  the Fourier components of $\bi p$ and $\bi u$,   
\be
{\hat{\bi p}}({\bi q})=\hspace{-1mm} \int\hspace{-1mm}
 d{\bi r} e^{-i{\bi q}\cdot{\bi r}}{\bi p}({\bi r}),~~
{\hat{\bi u}}({\bi q})=\hspace{-1mm} \int\hspace{-1mm}
 d{\bi r} e^{-i{\bi q}\cdot{\bi r}}{\bi u}({\bi r}),
\en 
where ${\bi q}= 2\pi ({n_x}/{L}, {n_y}/{L}, {n_z}/{H})$ 
with  $(n_x, n_y, n_z)$ being   integers for our 
$L\times L\times H$ system.  The wave number $q=|{\bi q}|$ 
is in the range $2\pi/H\le q<\pi/d$. It is convenient to 
divide ${\hat{\bi p}}({\bi q})$ into 
the longitudinal part  ${\hat{\bi p}}_\parallel=
({\hat{\bi p}}\cdot{\hat{\bi q}}) {\hat{\bi q}} $ 
and the  transverse part 
${\hat{\bi p}}_\perp={\hat{\bi p}}- 
{\hat{\bi p}}_\parallel$, where ${\hat q}= q^{-1}{\bi q}$.

We consider the contribution to 
${\cal  F}$ from   the Fourier components 
with ${\bi q}\neq {\bi 0}$,  retaining 
  the bilinear terms   and    the  third order ES  term. 
The resultant free energy is then divided 
into two parts as 
${\cal F}_{\rm inh}= {\cal F}_{\rm inh}^{\rm po}
+ {\cal F}_{\rm inh}^{\rm el}$, where  
\bea 
&& \hspace{-15mm}
{\cal F}_{\rm inh}^{\rm po} = \frac{1 }{2V}{\sum}_{{\bi q}\neq {\bi 0}}\Big[
4\pi {{\hat{\bi p}}_\parallel({\bi q})|^2 
+ (A+Cq^2)|{\hat{\bi p}}({\bi q})|^2}
\Big], \\
&&\hspace{-15mm} {\cal F}_{\rm inh}^{\rm el}= 
\frac{1 }{V}\hspace{-1mm}{\sum}_{{\bi q}\neq {\bi 0},\alpha,\beta}
\Big[ \frac{1}{2}q^2Z_{\alpha\beta}({\hat{\bi q}}) {\hat {u}}_\alpha({\bi q}) 
{\hat u}_\beta({\bi q})^* 
 -{\hat\Pi}_{\alpha\beta}^{\rm s}({\bi q})^*
iq_\beta  {\hat {u}}_\alpha({\bi q}) \Big].   
\ena  
In Eq.(24), the  Fourier component of $\bi E$ is set equal to 
$-4\pi {\hat{\bi p}}_\parallel$ from $\nabla\cdot{\bi E}=
 -\nabla^2\Phi= -4\pi \nabla\cdot{\bi p}$, where the surface effect 
is neglected.
In Eq.(25),  ${\sum}_\beta q^2Z_{\alpha\beta}({\hat{\bi q}}) {\hat {u}}_\beta({\bi q})$ 
is the Fourier component of $-{\sum}_\beta 
\nabla_\beta \sigma_{\alpha\beta}$, while 
${\hat\Pi}_{\alpha\beta}^{\rm s}({\bi q})$ is that of 
$\Pi^{\rm s}_{\alpha\beta}$.
The 
$\{Z_{\alpha\beta}\}$ is called 
 the Christoffel matrix \cite{Fedorov}: 
\bea 
&& Z_{\alpha\beta}({\hat{\bi q}})= 
  a_\alpha \delta_{\alpha\beta}+  (c_{12}+c_{44})
{\hat q}_\alpha{\hat q}_\beta,\\
&&a_\alpha({\hat{\bi q}}) =   
 c_{44}+ (c_{11}-c_{12}-2c_{44}){\hat q}_\alpha^2 . 
\ena 

From Eq.(24)  we obtain the polarization correlation 
\cite{Onuki2025},   
\bea
&&\hspace{-8mm}   { G}_{\alpha\beta}({\bi q}) =
 \av{{\hat p}_{\alpha} ({\bi q})
{\hat p}_{\beta}({\bi q})^*}\frac{1}{Vk_BT}
=\frac{{\hat q}_\alpha {\hat q}_\beta}{A+4\pi +Cq^2}+
\frac{\delta_{\alpha\beta}-
{\hat q}_\alpha {\hat q}_\beta}{A+Cq^2}~~({\bi q}\neq {\bi 0}). 
\ena 
The second term is the variance of     
${\hat{\bi p}}_\perp$, which grows 
for $q\xi\ls 1$ and $A\ll 1$ with  $\xi$ being  given in Eq.(7). 
Thus,  ${\hat{\bi p}}_\perp$ 
exhibits the critical fluctuations, while the 
longitudinal part  ${\hat{\bi p}}_\parallel$ is suppressed 
by the dipolar interaction. In the real space, 
the inverse Fourier  transforms of 
the two terms in Eq.(28) decay exponentially as 
$\exp[-r \sqrt{4\pi/C}]$ and 
$\exp[-r /\xi)$, respectively. 
 Aharony and Fisher\cite{Fisher} 
started with  Eq.(28)  to study the asymptotic critical behavior of 
 isotropic dipolar systems. For uniaxial non-piezoelectric systems 
such as  TGS,   
the $z$ component $p_z$ is the  single order parameter 
with the angle-dependent variance \cite{Leva1,Mina,Kri1,Kri2},  
$$\av{{\hat p}_{z} ({\bi q})
{\hat p}_{z}({\bi q})^*}/V= k_BT / (A+Cq^2+4\pi{\hat q}_z^2),$$ 
where $4\pi{\hat q}_z^2$ in the denominator 
arises from the dipolar interaction. 

From Eq.(3) the singular energy density   is given by 
$
e_{\rm sing}= f_{\rm B}- T\p f_{\rm B}/\p T= -a_1|{\bi p}|^2/2
$  
 in the lowest order in $\bi p$. Then, for $T>T_c$,  
we obtain the singular specific heat \cite{Str,Ono},  
\be 
C_{\rm sing}=\frac{\p}{\p T} \av{e_{\rm sing}}=
\frac{a_1^2}{4k_BT^2} \int \hspace{-1mm}d{\bi r}
\av{\delta |{\bi p}|^2({\bi r}) \delta |{\bi p}|^2({\bi 0})}
=  \frac{a_1^2}{4T} I_e A^{-1/2}, 
\en    
where we use  the decoupling approximation with Eq.(28). 
The coefficient $I_e$ is  defined by  
\be
  I_e=  {k_BT}/[{2\pi C^{3/2}}], 
\en 
where $C$ is the coefficient  in the gradient free energy. 
 Here, the factor  $a_1^2/4$ 
is extremely small, so the higher order terms 
in the free energy can well exceed the 
above $C_{\rm sing}$. Experimentally, the specific heat singularity 
in cubic BaTiO$_3$ has been  very small   \cite{Str,Ono,Iku}. 
To derive Eq.(29)  systematically we can introduce 
the fluctuating energy density variable $e$ with the 
coupling  $\propto e|{\bi p}|^2$ in $\cal F$ \cite{Ma,Onukib}.  
Note that the  four-body correlation in Eq.(29)   
 grows  nearly logarithmically as 
$\log A^{-1}$ in the asymptotic critical 
regime  \cite{Onukib,Ma}.

\vspace{2mm}
\noindent{\bf 3.3. Correlations of acoustic and electrostrictive 
displacements}\\
Accounting for  the ES coupling  in Eq.(11), we  
  divide the  Fourier  component  
${\hat {u}}_\alpha({\bi q})$  as  
\be 
{\hat {m}}_\alpha({\bi q})=- \frac{1}{q^2} {\sum}_{\beta,\gamma} 
Z^{\alpha\beta}({\hat{\bi q}}) iq_\gamma 
{\hat \Pi}_{\beta\gamma}^{\rm s}({\bi q}), 
~~~{\hat {w}}_\alpha({\bi q})= {\hat {u}}_\alpha({\bi q})
- {\hat {m}}_\alpha({\bi q}) ~~~({\bi q}\neq {\bi 0}). 
\en 
Here,   ${\hat {m}}_\alpha({\bi q})$ is the ES  part 
consisting of the Fourier components of the bilinear terms 
 $\propto p_\beta p_\gamma$, while 
  ${\hat {w}}_\alpha({\bi q})$ is the acoustic  part.   
 The matrix $\{Z^{\alpha\beta}({\hat{\bi q}})\}$ is the inverse  
of $\{Z_{\alpha\beta}({\hat{\bi q}})\}$ \cite{Onuki1989,Nam,Chen2}:       
\bea 
&&\hspace{-8mm}{Z^{\alpha\beta}({\hat{\bi q}})}=  
 \frac{ \delta_{\alpha\beta}}{a_\alpha({\hat{\bi q}})}
-\frac{(c_{12}+c_{44})
{{\hat q}_\alpha{\hat q}_\beta}}{{a_\alpha({\hat{\bi q}}) 
a_\beta({\hat{\bi q}})(1+b({\hat{\bi q}}))}},\\
&& \hspace{-8mm}
b({\hat{\bi q}})= (c_{12}+c_{44}){\sum}_\gamma 
{{\hat q}_\gamma^2}/{a_\gamma({\hat{\bi q}})}, 
\ena 
where $b({\hat{\bi q}})= 3.30$ for ${\hat{\bi q}}$ along $[001]$ from Table 1. 
In Sec.IV,  we shall see that ${\hat {m}}_\alpha$ 
and ${\hat {w}}_\alpha$  evolve in time very  differently, 
resulting in the  central and acoustic peaks, respectively.  

In the real space, we have  $m_\alpha({\bi r})=
u_\alpha({\bi r})-w_\alpha({\bi r})=
{\sum}_{{\bi q}\neq{\bi 0}}e^{i{\bi q}\cdot{\bi r}}
{\hat {m}}_\alpha({\bi q})/V$. 
If $\bi r$ is in the bulk region, we obtain the integral expression, 
\be
m_\alpha({\bi r})= -{\sum}_{\beta,\gamma}
\int d{\bi r}' 
{\cal K}_{\alpha\beta}({\bi r}-{\bi r}') \nabla'_\gamma 
\Pi^{\rm s}_{\beta\gamma}({\bi r}').
\en 
Here,     ${\cal K}_{\alpha\beta} ({\bi r}) $
are  the  Green functions in the limit $V\to \infty$  given by   
\be 
 {\cal K}_{\alpha\beta} ({{\bi r}}) = \frac{1}{(2\pi)^{3}}
 \int d{\bi q}e^{i{\bi q}\cdot{\bi r}}\frac{1}{q^2} 
Z^{\alpha\beta}({\hat{\bi q}}) =
 \frac{1}{8\pi^2r} 
\int d\theta d\varphi\sin\theta ~ 
Z^{\alpha\beta}({\hat{\bi q}})\delta({\hat{\bi q}}\cdot {\hat{\bi r}}) . 
\en 
where $\int  d\theta d\varphi\sin\theta (\cdots)$ 
is  the  angle integration  of  ${\hat{\bi q}}$.  
The   delta function $ 
\delta({\hat{\bi q}}\cdot {\hat{\bi r}})$ 
 follows from   $Z^{\alpha\beta}(-{\hat{\bi q}})=
Z^{\alpha\beta}({\hat{\bi q}})$, 
where  ${\hat{\bi r}}=r^{-1} {\bi r}$. 
Thus,   $r{\cal K}_{\alpha\beta} ({\bi r})$ 
depends only on  ${\hat{\bi r}}$. If 
${\hat{\bi r}}= (0,0,1)$, 
we have $ r{\cal K}_{\alpha\beta} ({\hat{\bi r}}) = 
 \int_0^{2\pi} d{\varphi } Z^{\alpha\beta}({\hat{\bi q}})/8\pi^2$ with ${\hat{\bi q}}=(\cos\varphi,\sin\varphi,0) $. 
We note  that the elastic field around an impurity 
can be expressed  in terms of ${\cal K}_{\alpha\beta} ({\bi r})$ 
in cubic solids.

With the division of the displacement 
in Eq.(31),  ${\cal F}_{\rm inh}^{\rm el}$ in  Eq.(25) 
is rewritten as 
\be
{\cal F}_{\rm inh}^{\rm el}= \frac{1}{2V} 
{\sum}_{{\bi q}\neq {\bi 0},\alpha,\beta}
q^2Z_{\alpha\beta}({\hat{\bi q}}){\hat {w}}_\alpha({\bi q})
{\hat w}_\beta({\bi q})^* 
 + \Delta{\cal F}^{\rm el}_{\rm inh}(\{{\bi p}\}), 
\en
where   $\Delta{\cal F}^{\rm el}_{\rm inh}$ 
depends only on $\bi p$ as will be shown  in Eq.(51). 
The variables ${\hat {w}}_\alpha({\bi q})$ 
and ${\hat {p}}_\alpha({\bi q})$ are 
  statistically independent of each other. Thus, 
  the displacement correlation, written as  
$U_{\alpha\beta}({\bi q})$,  is divided  into 
those  of ${\hat w}_\alpha$ and   
  ${\hat m}_\alpha$ as  
\bea 
&&\hspace{-12mm}U_{\alpha\beta}({\bi q})=
\av{{\hat u}_{\alpha} ({\bi q})
{\hat u}_{\beta}({\bi q})^*}/V=
\av{{\hat w}_{\alpha} ({\bi q})
{\hat w}_{\beta}({\bi q})^*}/V+ 
\av{{\hat m}_{\alpha} ({\bi q})
{\hat u}_{\beta}({\bi q})^*}/V\nonumber\\
&&=  
 k_BT Z^{\alpha\beta}({\hat{\bi q}})/q^2+ R_{\alpha\beta}({\bi q}). 
\ena 
The ES correlation $R_{\alpha\beta}({\bi q})$   is 
 written in terms of 
 the stress correlation $\Psi_{\alpha\beta}({\bi q})$ as 
\bea 
&& R_{\alpha\beta}({\bi q})=\frac{1}{q^2}k_BT 
{\sum}_{\gamma,\nu} 
Z^{\alpha\gamma}({\hat{\bi q}})Z^{\beta\nu}({\hat{\bi q}})
\Psi_{\gamma\nu}({\bi q}). \\
&&\Psi_{\alpha\beta}({\bi q})=\hspace{-1mm} \frac{1}{k_BT}{\sum}_{\gamma,\nu} 
{\hat q}_\gamma {\hat q}_\nu \int \hspace{-1mm}
d{\bi r} e^{-i{\bi q}\cdot{\bi r}} 
\av{\delta\Pi_{\alpha\gamma}^{\rm s}({\bi r})
\delta\Pi_{\beta\nu}^{\rm s}({\bi 0})},
\ena 
where $\delta \Pi_{\alpha\gamma}^{\rm s}= 
\Pi_{\alpha\gamma}^{\rm s} -\av{\Pi_{\alpha\gamma}^{\rm s}}$.  
Hereafter, to avoid cumbersome notation, we will write 
 $Z_{\alpha\beta}({\hat{\bi q}})$, $a_\alpha({\hat{\bi q}})$, 
 $Z^{\alpha\beta}({\hat{\bi q}})$, and $b({\hat{\bi q}})$ 
 as  $Z_{\alpha\beta}$, $a_\alpha$, 
 $Z^{\alpha\beta}$, and $b$ 
  unless confusion may occur. 

In this paper, we use $\Psi_{\alpha\beta}({\bi q})$ in 
  the limit  $q\to 0$ assuming    $q\xi\ll 1$. Then, in Eq.(39), 
$e^{-i{\bi q}\cdot{\bi r}}$  can be set equal to 1 
because the integrand 
decays  exponentially ($\propto e^{-2r/\xi}$). However, 
  $\Psi_{\alpha\beta}({\bi q})$  still depends on 
 ${\hat{\bi q}}$ because of the factor ${\hat q}_\gamma {\hat q}_\nu $.  
In  Appendix C,   we will find that 
 $\Psi_{\alpha\beta}({\hat{\bi q}})={\lim}_{q\to 0}
\Psi_{\alpha\beta}({{\bi q}})$ grows as $A^{-1/2}$ 
in the following form, 
\bea 
&&\hspace{-10mm}
\Psi_{\alpha\beta}({\hat{\bi q}}) =  
\Big[\Big(M_{44} + (M_{11}-M_{12}-2M_{44})
{\hat q}_\alpha^2\Big) \delta_{\alpha\beta}
+(M_{12}+M_{44}){\hat q}_\alpha{\hat q}_\beta\Big]I_eA^{-1/2}.   
\ena 
In terms of  $g_{ij}$  in Eq.(12), 
the dimensionless coefficients   $M_{ij}$   are given by   
\be 
M_{11} -M_{12}= {7} (g_{11}- g_{12})^2/30,~~
M_{11}+2M_{12}= (g_{11}+ 2g_{12})^2/3,~~
M_{44} = {7} g_{44}^2/60. 
\en  
In  Table 1 we give   $M_{ij}$  for BaTiO$_3$, 
where  $M_{11}-M_{12}$ 
is the largest  combination and $M_{44}$ 
is very small. Thus, $\Psi_{\alpha\beta}({\hat{\bi q}}) 
\cong (M_{11}-M_{12})
{\hat q}_\alpha^2 \delta_{\alpha\beta}I_eA^{-1/2}$. 
 From the cubic symmetry,   $Z_{\alpha\beta}({\bi q})$ in Eq.(26) 
and  $\Psi_{\alpha\beta}({\hat{\bi q}})  $ 
in Eq.(40)  assume  the same 
 tensorial  form.

 \begin{figure}[t]
\includegraphics[scale=1.2]{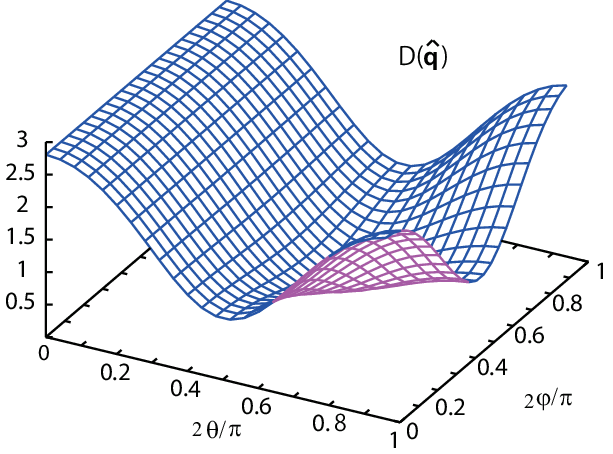}
\caption{\protect  Anisotropy factor  $D({\hat{\bi q}})$ 
in Eqs.(42) and (43) in  the density correlation 
 for cubic BaTiO$_3$ as a function of $(\theta,\varphi)$ 
with   $ {\hat{\bi q}}= (\sin\theta\cos\varphi, \sin\theta\sin\varphi, 
\cos\theta)$. Its  maximum  is 2.8, which  is attained at  
 $\theta=0$ ($[001]$),  
  $(\theta,\varphi)=(\pi/2,0)$ ($[100]$), 
and $(\theta, \varphi)=(\pi/2,\pi/2)$ ($[010]$). 
Its minimum is 0.44 for $(\theta,\varphi)=(\pi/4,\pi/4)$ ($[111]$). Use is made of $c_{ij}$ and $M_{ij}$ in Table 1.}
\end{figure}

\vspace{2mm}
\noindent
{\bf 3.4.  Density correlation and entropy fluctuation    }\\ 
From Eq.(10) we have 
${\hat\rho}({\bi q})=\int d{\bi r}e^{-i{\bi q}\cdot{\bi r}}
\delta\rho({\bi r})=-\rho i{\bi q}\cdot{\hat{\bi u}}({\bi q})$. 
Then,   Eq.(37) gives     
\bea 
&&\hspace{-11mm} I_D({\bi q})={
{\av{|{\hat \rho} ({\bi q})|^2}}}/{{V}\rho^2} 
={\sum}_{\alpha,\beta} {q}_\alpha { q}_\beta 
U_{\alpha\beta}({\bi q})\nonumber\\
&& 
=\frac{k_BTb({\hat{\bi q}})}{(c_{12}+c_{44}){(1+b({\hat{\bi q}}))} }
 + \frac{k_BT I_e  }{(c_{12}+c_{44})^2}D({\hat{\bi q}}) A^{-1/2}.  
\ena 
In the second line, 
   the  first term is the 
variance of  $- i{\bi q}\cdot{\hat{\bi w}}({\bi q})$ 
and the second  term is  that of  
$- i{\bi q}\cdot{\hat{\bi m}}({\bi q})$. 
The first term becomes $k_BT/c_{11}$ 
for ${\hat{\bi q}}$ along $[001]$.  
From  ${\sum}_\gamma Z^{\alpha\gamma}
{\hat q}_\gamma= {\hat q}_\alpha/[a_\alpha (1+b)]$.  
we express the factor   $D({\hat{\bi q}})$  in the  second term  as 
\bea 
&&\hspace{-10mm} D({\hat{\bi q}})=
\frac{(c_{12}+c_{44})^2}{k_BTI_e}{\sum}_{\alpha,\beta} 
{\hat{q}}_\alpha {\hat{q}}_\beta R_{\alpha\beta}
= \frac{b^2}{(1+b)^2}(M_{12}+ M_{44}) \nonumber\\
&&+ 
\frac{(c_{12}+ c_{44})^2}{(1+b)^2} 
{\sum}_\gamma    \frac{{\hat{q}}_\gamma^2}{a_\gamma^2}\Big[ 
M_{44}+ (M_{11}-M_{12}-2M_{44}){\hat{q}}_\gamma^2\Big], 
\ena 
where $b({\hat{\bi q}})$ is written as $b$. 
In Fig.1, we display $D({\hat{\bi q}})$ for cubic BaTiO$_3$. 
Its   maximum is $D[001]=2.8$  
 along  $[001]$ and  its  minimum is $D[111]=0.44$  
 along  $[111]$, while it  is $D[110]=0.94$ along $[011]$. 
The ratio of the second term to the first one in Eq.(42) 
is about $0.4\times 10^{-2}A^{-1/2}$ along $[001]$.  which increases 
 up to  0.1 at $T=T_c$ for  BaTiO$_3$.

In $I_D({\bi q})$ in Eq.(42) 
the contribution from the entropy fluctuation $\delta s({\bi r})$ 
(per unit mass)  is omitted, which 
 is crucial for compressible  fluids \cite{water,Martin,Onukib}. 
Note that   the isobaric part of the  density 
deviation  $(\p \rho/\p s)_p \delta s$ 
contributes $k_B T^2\alpha_P^2/ C_P$  
  to $I_D({\bi q})$, which is the integral 
intensity of the Rayleigh scattering. Here, 
the first term in Eq.(42) is of order 
$k_BT/c_{11}\sim k_BT(\p \rho/\p P)_s/\rho$ from Eq.(18). 
Thus, in Eq.(42), the omitted   entropy 
contribution  is  smaller than the first term 
   by a factor of the Landau-Placzek ratio \cite{Landau-e}, 
\be 
\gamma-1= 
C_P/C_V-1= (c_{11}+2c_{12})T\alpha_P^2/3C_P,  
\en 
where $\gamma-1 \sim 2.5\times 10^{-3}$ 
for BaTiO$_3$ with  $C_P$ and $C_V$ 
being     the isobaric and isochoric  
specific heats per unit volume, respectively. 
We also find   $\gamma-1 \sim 10^{-2}$ 
for KDP and $\gamma-1\sim 10^{-3}$ for TGS. 
However,  some  experimental groups   \cite{Ly1,Cum2,Y2} 
ascribed the origin of the central peak 
near the ferroelectric transition  
to  the Rayleigh scattering.

\vspace{2mm}
\noindent{\bf 3.5. Elastic free energy in terms of  
 renormalized elastic moduli $C_{ij}$}\\
Let us  integrate  the statistical 
distribution over  $\bi p$ at fixed $\bi u$   as 
\be 
\int {d}\{{\bi p}\} \exp[-{\cal F}(\{{\bi p}\},\{{\bi u}\})/k_BT]
= {\rm const.}
\exp\Big[-\frac{1}{2V}{\sum}_{{\bi q}\neq {\bi 0},\alpha,\beta}
 U^{\alpha\beta}({\bi q})  {\hat {u}}_\alpha({\bi q})
{\hat u}_\beta({\bi q})^*\Big],
\en  
where $\int {d}\{{\bi p}\}(\cdots)$ denotes the functional
 integration and we consider the distribution of ${\hat {u}}_\alpha({\bi q})$ 
  with ${\hat{\bi q}}\neq {\bi 0}$. 
Here,  $\{ U^{\alpha\beta}({\bi q})\}$ is 
  the  inverse matrix of $\{ U_{\alpha\beta}({\bi q})\}$ 
in Eq.(37). We can then define  the renormalized, static 
  elastic moduli  $C_{ij}$ by 
\bea 
&&\hspace{-10mm}
 U^{\alpha\beta}({\bi q}) =\Big[ 
\Big(C_{44}+ (C_{11}-C_{12}-2C_{44}){\hat q}_\alpha^2\Big) 
 \delta_{\alpha\beta}+  (C_{12}+C_{44})
{\hat q}_\alpha{\hat q}_\beta\Big]q^2/k_BT,
\ena 
where    $c_{ij}$   
in $Z_{\alpha\beta}$ in Eq.(26)   are replaced by $C_{ij}$ in Eq.(46). 
Since   Eqs.(37) and (38) give 
 $U_{\alpha\beta}q^2/k_BT= Z^{\alpha\beta} 
+ {\sum}_{\gamma,\nu} Z^{\alpha\gamma} \Psi_{\gamma\nu}Z^{\nu\beta}$,  
we can expand 
  $ U^{\alpha\beta}({\bi q})$ 
 with respect to  $\Psi_{\alpha\beta}({\bi q})$ as 
\be 
 U^{\alpha\beta}({\bi q})
= [Z_{\alpha\beta}- \Psi_{\alpha\beta}({\bi q})+\cdots]
q^2/k_BT.
\en  
Then,  to linear order in $\Psi_{\alpha\beta}({\bi q})$, 
  Eqs.(26), (27), (40), (46), and (47)  give   
\be 
C_{ij} = c_{ij}-  M_{ij}I_eA^{-1/2}. 
\en 
The renormalized 
frequency-dependent elastic moduli will be given in Eq.(78),

From  Eq.(48) we find    
$ \Delta C_{11}/c_{11} 
=1-C_{11}/c_{11}= 
2.2\times 10^{-4}(d/\sqrt{C})^{3}A^{-1/2}$  
with $d=4~{\rm \AA}$, 
where we  use   $c_{11}$ and $M_{11}$ in Table 1. 
Experimentally, Kashida {\it et al}. \cite{Kashida}  found 
\be 
\Delta C_{11}/c_{11}=0.032(T/T_0-1)^{-1.2}= 
0.43\times 10^{-2}A^{-1/2}. 
\en  
We equate our  theoretical expression    to 
the above experimental one to  determine 
the coefficient $C$ in  the gradient  free energy as   
\be 
\sqrt{C}=1.5 ~{\rm \AA}.  
\en 
Then, $\xi= 1.5A^{-1/2}~{\rm \AA}=
11 (T/T_0-1)^{-1/2} ~{\rm \AA}$ from Eq.(6a). 

Li {\rm et al.}\cite{Ko1} determined 
 all $\Delta C_{ij}$ in cubic BaTiO$_3$ by Brillouin scattering,  
where  $\Delta C_{11}/c_{11}\sim 0.1$ 
 and   $\Delta C_{44}/c_{44}\sim  0.02$ at $T=T_c$.  	
Their  findings agree  with  Eq.(49) 
and  $M_{ij}$ in Table 1. 
Thus, these previous results \cite{Ko1,Kashida,Chen1}  
 justify  the expansion of $U^{\alpha\beta}({\bi q})$ 
with respect to $\Psi_{\alpha\beta}({\bi q})$ in Eq.(47).

\vspace{2mm}
\noindent{\bf 3.6. Free energy of $\bi p$ 
after removal of $\bi u$}\\  
The  free energy contribution 
from the inhomogeneous ES displacement ${\hat{\bi m}}({\bi q})$ becomes
\bea 
&&\hspace{-12mm} \Delta{\cal F}^{\rm el}_{\rm inh}=- \frac{1}{2V}
{\sum}_{{\bi q}\neq {\bi 0},\alpha,\beta,\gamma,\nu}
Z^{\alpha\beta}({\hat{\bi q}}) 
 {\hat q}_\gamma{\hat q}_\nu {\hat{\Pi}}_{\alpha\gamma}^{\rm s}({\bi q})
{\hat{\Pi}}_{\beta\nu}^{\rm s}({\bi q})^*  
\nonumber\\ 
&&\hspace{-0mm} 
= -\frac{1}{2}\int d{\bi r}\int d{\bi r}'~ 
{\sum}_{\alpha,\gamma,\beta,\nu} 
{\cal K}_{\alpha\beta} ({\bi r}-{\bi r}') 
\nabla_\gamma \Pi^{\rm s}_{\alpha\gamma}({\bi r}) \cdot 
\nabla'_\nu \Pi^{\rm s}_{\beta\nu}({\bi r}'), 
\ena 
which is the second term in Eq.(36). 
In the second line,  we use 
 ${\cal K}_{\alpha\beta} ({\bi r}) $  in Eq.(35) 
neglecting the surface effect. After removal of $\bi u$  we obtain the 
 free energy of ${\bi p}$  in the form, 
\be 
{\cal F}_R(\{ {\bi p}\})= \int d{\bi r}\Big[\frac{1}{8\pi}|{\bi E}|^2  
+f_{\rm B}\Big]+  \Delta{\cal F}^{\rm el}_{\rm inh}
+ {\Delta{\cal F}_{\rm hom}^{\rm el}}. 
\en 
Here,  ${\Delta{\cal F}_{\rm hom}^{\rm el}}$ arises 
from  homogeneous strains 
in the stress-free boundary 
condition (or at a fixed ambient pressure)  
in unclamped samples, as will be derived in Appendix A.

We make comments. (i)  For   
 cubic perovskites,  Nambu and Sagara \cite{Nam} solved the dynamic 
equation $\tau_0 \p {\bi p}/\p t= -
\delta {\cal F}_R /\delta {\bi p}$  (see Eq.(53)) 
after loweing $T$ below $T_0$ in two dimensions,  
In ${\cal F}_R$ they included 
  $\Delta{\cal F}^{\rm el}_{\rm inh}$ 
and  found its relevance  in phase ordering. 
However,  they set $ {\Delta{\cal F}_{\rm hom}^{\rm el}}=0$ 
in  the periodic boundary condition 
and ${\bi E}={\bi 0}$ without  the dipolar interaction. 
 (ii) In phase-separating  binary alloys forming a cubic 
crystal,  the composition $\phi({\bi r})$ is  the order parameter 
  with the Vegard  coupling 
 $g \phi \nabla\cdot {\bi u}$  in the free energy, 
where  $g$ is a constant \cite{Nishi,Onukib}. 
In this case,   we can use Eq.(51) to obtain  
  $\Delta{\cal F}^{\rm el}_{\rm inh} 
= {\sum}_{{\bi q}\neq {\bi 0}}
r_{\rm el}({\hat{\bi q}}) |{\hat \phi}({\bi q})|^2/2V$, 
where ${\hat \phi}({\bi q})$ is 
 the Fourier component of $\phi({\bi r})$ and 
$$r_{\rm el}({\hat{\bi q}})
=- g^2 {\sum}_{\alpha,\beta} Z^{\alpha\beta}({\hat{\bi q}})
{\hat{q}}_\alpha {\hat{q}}_\beta= -g^2
b({\hat{\bi q}})/[(1+b({\hat{\bi q}}))(c_{12}+ c_{44})].$$ 
The    ${\hat{\bi q}}$-dependence of 
$\tau_{\rm el}({\hat{\bi q}})$     
determines the domain  structure  in phase ordering.

\vspace{2mm}
\noindent{\bf 4. Dynamics of ${\bi p}$ and $\bi u$ }\\
\noindent{\bf 4.1. Polarization time-correlation and 
frequency-dependent dielectric constant}\\ 
 Now, we examine  dynamics of  
${\bi p}({\bi r},t)$ and ${\bi u}({\bi r},t)$,  which  
 depend  on space and time.  Starting with  Eq.(19)  we set up  
the relaxation equation  in the linear regime 
 \cite{Hubb,Khala,Onuki2025,Onukib,Chen2,Nam}, 
\be 
\tau_0 \frac{ \p }{\p t} {\bi p}=  - 
\Big(\frac{\delta{\cal F}}{\delta {\bi p}}\Big)_{{\bi u}}
= {\bi E}+C\nabla^2{\bi p}- A{\bi p}, 
\en  
where $\tau_0$ is a microscopic time (to be estimated below Eq.(71)) 
and the nonlinear terms in the functional derivative in  Eq.(19) 
are omitted. In Eq.(53) we also neglect 
 the rapidly relaxing 
background polarization (such as the electronic one) 
 \cite{Shirane,Vogt,Pono}. 

We set 
the Fourier component of  ${\bi E}$ 
 equal to $-4\pi{\hat{\bi p}}_\parallel$ for ${\bi q}\neq {\bi 0}$ 
from $\nabla\cdot{\bi D}=0$ in Eq.(53) to examine the bulk correlations. Then, 
the longitudinal part ${\hat{\bi p}}_\parallel$ 
and the transverse part ${\hat{\bi p}}_\perp$ 
of the Fourier component ${\hat{\bi p}}$ obey
 \be 
 \frac{\p }{\p t}{\hat{\bi p}}_\parallel
 =- \Gamma_\parallel(q){\hat{\bi p}}_\parallel, ~~~
  \frac{\p }{\p t}{\hat{\bi p}}_\perp
 =- \Gamma_\perp(q){\hat{\bi p}}_\perp ~~~~({\bi q}\neq{\bi 0}),
\nonumber
\en 
where  the longitudinal and transverse 
decay  rates are given by 
\be 
\Gamma_\parallel(q)=   (A+4\pi +Cq^2)/\tau_0,~~~~
 \Gamma_\perp(q)=  (A+Cq^2)/\tau_0.
\en 
For small $q$, we have 
 $\Gamma_\perp(q)\ll \Gamma_\parallel(q)$, so  
${\hat{\bi p}}_\perp$ 
relaxes  slowly  on the time scale of $\tau_D$ with   
\be 
\tau_D={\lim}_{q\to 0} 
\Gamma_\perp(q)^{-1} =\tau_0/A.     
\en  
Here, $\tau_D$ is the Debye relation time 
and will be estimated  for  BaTiO$_3$ 
(see Eqs.(59) and (71)). 
On the other hand, $\bi u$ obeys   
  the dynamic elasticity equation,  
\be 
\rho \frac{\p^2}{ \p t^2} u_\alpha
=-\Big(\frac{\delta {\cal F}}{\delta u_\alpha}\Big)_{{\bi p}}
 = {\sum}_\beta \nabla_\beta \sigma_{\alpha\beta}
-{\sum}_\beta \nabla_\beta \Pi^{\rm s}_{\alpha\beta},  
\en  
where  $\sigma_{\alpha\beta}$ and $\Pi^{\rm s}_{\alpha\beta}$ 
are given in Eqs.(9) and (12) and the last term in Eq.(20) 
is neglected. The dissipatin of $\bi u$ arises from the  
thermal fluctuations 
of the ES force density $-{\sum}_\beta\nabla_\beta 
\Pi^{\rm s}_{\alpha\beta}$.

To describe the thermal fluctuations 
 of $\bi  p$,  we should treat  Eq.(53) 
as a  Langevin equation to  by adding 
a random noise term related to the 
kinetic coefficient   $1/\tau_0$ \cite{Onukib,Ma}.  
Then, 
we can calculate the polarization time-correlation function 
${\hat G}_{\alpha\beta}({\bi q},t)$, which 
 is divided into the  longitudinal and transverse parts as    
\bea 
&&\hspace{-10mm}
{\hat G}_{\alpha\beta}({\bi q},t)=
{ \av{{\hat{ p}}_\alpha ({\bi q},t)
{\hat{p}}_\beta({\bi q},0)^*}}/{Vk_BT}
={\hat G}_{\parallel}({\bi q},t){\hat{ q}}_\alpha {\hat{ q}}_\beta
+{\hat G}_{\perp}({\bi q},t)(\delta_{\alpha\beta}-{\hat{ q}}_\alpha 
{\hat{ q}}_\beta) , \nonumber\\ 
&& \hspace{-10mm} 
{\hat G}_{\parallel}({\bi q},t)=\hspace{-1mm}  
e^{-t\Gamma_\parallel(q)}/(A+4\pi +Cq^2),~~~~ 
{\hat G}_{\perp}({\bi q},t)
=\hspace{-1mm} e^{-t\Gamma_\perp(q)}/(A +Cq^2),
\ena 
From Eq.(54)  ${\hat G}_{\perp}({\bi q},t)$ decays slowly  
and   ${\hat G}_\parallel({\bi q},t)$  rapidly at small $q$. 
We neglect the third order ES coupling in Eqs.(53)-(57). 
 For highly polar fluids,  Ladanyi and  Skaf  \cite{Lada} 
confirmed Eq.(57)  at small $q$ 
using   molecular dynamics simulation.

We also present  the dynamic equation for  the sample average 
${\bar{p}}_z(t)=\int d{\bi r} { p_z}({\bi r},t)/V$ of the $z$ 
component $p_z$. In the geometry of  parallel metal plates, 
it reads      
\be 
 \frac{\p }{\p t}{\bar{p}}_z =-\frac{1}{\tau_D}
 {\bar{p}}_z + \frac{1}{\tau_0}E_a
= -\frac{1}{\tau_D}
 {\bar{p}}_z + \frac{4\pi}{\tau_0}({\bar\sigma}_0-{\bar{p}}_z). 
\en 
Here, ${\bar p}_z$ relaxes with the rate $1/\tau_D$ 
at fixed  $E_a= \Phi_a/H$ and  with $1/\tau_D+4\pi=(4\pi+A)/\tau_0$ 
at fixed ${\bar\sigma}_0=Q_0/L^2$  
from $E_a= 4\pi ({\bar\sigma}_0- {\bar p}_z)$, 
where the surface potential drops are neglected.  
 This dramatic difference of 
the relaxation rates of ${\bar{p}}_z $ is   
known for polar fluids \cite{Sprik,Onuki2025}. 

We   consider the thermal average $\av{{\bar p}_z(t)}$ induced 
by a   small oscillatory $E_a=\Phi_a/H$ along the $z$ axis, 
which  depend on $t$ as  $ e^{i\omega t}$ 
in the complex number representation. 
Then, for large systems, Eq.(58) gives  
  the frequency-dependent dielectric  constant 
\cite{Hatta,Hill,Vogt,Pono},  
\be 
\ep^*(\omega)=4\pi \av{{\bar{p}}_z(t)}/E_a= 
\ep /(1+i\omega\tau_D) .
\en 
Thus,   $\tau_D$ in Eq.(55) is the Debye relaxation time. 
 Furthermore, for thin films with 
thickness $H$,  Eq.(21) is generalized to give 
 the effective frequency-dependent dielectric  constant  \cite{Onuki2025},  
\be 
\ep_{\rm eff}^*(\omega)=
[ 1/\ep^*(\omega)+ 2\ell_s/H]^{-1}
=\ep_{\rm eff}/(1+i\omega\tau_D'), 
\en 
where the relaxation time is 
decreased as $\tau_D'=\tau_D/(1+ 2\ep\ell_s/H)$ 
due to finite $H$.

\vspace{2mm}
\noindent{\bf 4.2. Time-correlation 
of electrostrictive displacement}\\ 
We first consider  the ES displacement ${\hat{ m}}_\alpha({\bi q},t)$ 
 in  Eq.(31), whose time-evolution 
is governed by Eq.(53). 
The time-correlation function  
of  ${\hat{\bi m}}({\bi q},t)$   and its FL transform 
are written  as  
\bea
&&{\hat R}_{\alpha\beta}({\bi q},t)=
{ \av{{\hat{ m}}_\alpha ({\bi q},t)
{\hat{m}}_\beta({\bi q},0)^*}}/{V}
=\frac{k_BT}{q^2} {\sum}_{\gamma,\nu} 
Z^{\alpha\gamma}Z^{\beta\nu}{\hat\Psi}_{\gamma\nu}({\hat{\bi q}},t).
\\
&& 
R^*_{\alpha\beta}({\bi q},\omega)=
\int_0^\infty dt  e^{-i\omega t} 
{\hat R}_{\alpha\beta}({\bi q},t)
=\frac{k_BT}{q^2} {\sum}_{\gamma,\nu} 
Z^{\alpha\gamma}Z^{\beta\nu}\zeta^*_{\gamma\nu}(\omega).
\ena
Here,   ${{\hat\Psi}}_{\alpha\beta}({\hat{\bi q}},t) $ 
is the stress time-correlation function 
and  ${{\zeta}}^*_{\alpha\beta}({\hat{\bi q}}, \omega)$ is its FL transform, 
\bea 
&& {{\hat\Psi}}_{\alpha\beta}({\hat{\bi q}},t)=\frac{1}{k_BT}
\sum_{\gamma,\nu}{{\hat q}_\gamma{\hat q}_\nu} 
\int \hspace{-1mm}
d{\bi r} \av{\delta\Pi_{\alpha\gamma}^{\rm s}({\bi r},t)
 \delta \Pi_{\beta\nu}^{\rm s}({\bi 0},0)} ,\\
&& 
{{\zeta}}^*_{\alpha\beta}({\hat{\bi q}}, \omega)=
\int_0^\infty\hspace{-1mm} dt  e^{-i\omega t} 
{{\hat\Psi}}_{\alpha\beta}({\hat{\bi q}},t),
\ena 
where $\delta\Pi_{\alpha\gamma}^{\rm s}=
\Pi_{\alpha\gamma}^{\rm s}- \av{\Pi_{\alpha\gamma}^{\rm s}}$.  
From Eq.(37) we have 
$ {{\hat\Psi}}_{\alpha\beta}({\hat{\bi q}},0)=
 {{\Psi}}_{\alpha\beta}({\hat{\bi q}})$ 
and ${{\zeta}}^*_{\alpha\beta}({\hat{\bi q}}, \omega)
\to  {{\Psi}}_{\alpha\beta}({\hat{\bi q}})/i\omega$ 
as $\omega\to\infty$.  
 We  call  $\zeta^*_{\alpha\beta}(\omega)$  
 the  frequency-dependent viscosity  
 in solids \cite{Landau-ela,Zacc}.  
Though  $Z^{\alpha\beta}({\hat{\bi q}})$, 
$\Psi_{\alpha\beta}({\hat{\bi q}})$, 
and ${{\zeta}}^*_{\alpha\beta}({\hat{\bi q}}, \omega)$ 
  depend on  ${\hat{\bi q}}$  even in the limit  $q\xi\ll 1$,  
 their ${\hat{\bi q}}$-dependence will be suppressed 
in many relations to follow.

 \begin{figure}[t]
\includegraphics[scale=0.6]{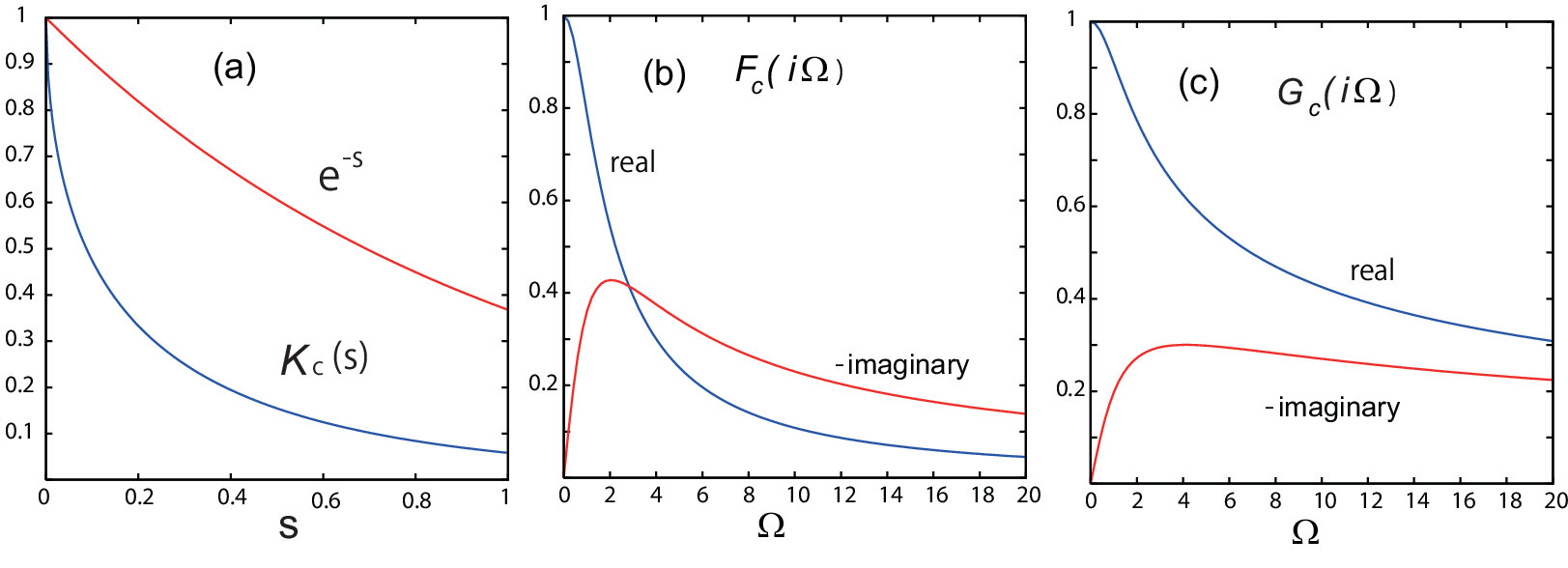}
\caption{\protect  
(a) Scaling  function 
$K_c(s)$ (blue line) in Eq.(68)  with 
$s=2t/\tau_D$, which has a cusp at $s=0$ 
and decays faster than $e^{-s}$ (red line). 
 (b) Real part and imaginary 
parts of the scaling function $F_c(i\Omega)$ in Eq.(69)  
with $\Omega=\omega \tau_D/2$, which decay slowly for $\Omega\gs 4$. 
(c) Real part and imaginary 
parts of the scaling function $G_c(i\Omega)$ in Eq.(77), 
which also decay slowly for $\Omega\gs 4$.
 }
\end{figure}
In    Appendix C, we shall see that 
the above correlation functions depend on $t$ and $\omega$  as 
 \bea 
&&\hspace{-8mm}
{{\hat\Psi}}_{\alpha\beta}({\hat{\bi q}},t)=
\Psi_{\alpha\beta}({\hat{\bi q}})
  K_c(2t/\tau_D) ,~~~~
 {{\zeta}}^*_{\alpha\beta}({\hat{\bi q}}, \omega)=
\Psi_{\alpha\beta}({\hat{\bi q}}) 
F_c(i\Omega){\tau_D}/{8} ,\\
&&\hspace{-8mm}
{{\hat R}}_{\alpha\beta}({\hat{\bi q}},t)=
R_{\alpha\beta}({\hat{\bi q}})
  K_c(2t/\tau_D) ,~~~
 R^*_{\alpha\beta}({\hat{\bi q}},\omega)=
R_{\alpha\beta}({\hat{\bi q}})
  F_c(i\Omega)\tau_D/8 ,
\ena 
where the static correlations $R_{\alpha\beta}({\hat{\bi q}})$ in Eq.(38) 
and $\Psi_{\alpha\beta}({\hat{\bi q}})$
in Eqs.(39) and (40) grow as $A^{-1/2}$. 
Here, $t$ is scaled by $\tau_D/2$ and $\omega$ by $2/\tau_D$. 
Thus,   we introduce  the  scaled frequency,    
\be 
\Omega= \omega\tau_D/2= \omega\tau_0/2A. 
\en  
The  scaling functions $K_c(s) $ and $F_c(x)$ in Eqs.(65) 
and (66) are defined  by 
\bea 
&& K_c(s) =\frac{4}{\pi} \int _0^\infty dx 
\frac{x^2\exp[-s-sx^2]}{(1+x^2)^2}
=\sqrt{8/\pi}D_{-3}(\sqrt{2s}),\\
&&F_c(x)= 4\hspace{-1mm}\int_0^\infty \hspace{-2mm}ds e^{-xs}
 K_c(s)= \frac{8}{x^2}\Big[1+ \frac{x}{2}- \sqrt{1+x}\Big], 
\ena 
where   $K_c(s)$ is proportional to $ 
\int d{\bi q}{\hat G}_\perp(q,t)^2$,  
$D_{-3}(x)$  is the Weber  function,  and the 
   Laplace transform of $K_c(s)$  is   $F_c(x)/4$. 
In Fig.2, we plot $K_c(s)$ and $F_c(i\Omega)$. 
Here,   $K_c(s)\cong 1-4\sqrt{s/\pi}$ for $0<s\ll 1$ and  
$ K_c(s)\cong \sqrt{\pi}e^{-s}/s^{3/2}$ for $s\gg 1$, 
 while $F_c(i\Omega)\cong 1-i\Omega/2$ for $\Omega\ll 1$  
and $F_c(i\Omega)\cong  4/i\Omega-8/(i\Omega)^{3/2}$ for $\Omega\gg 1$. 
The $K_c(s)$ has a cusp at $s\cong 0$ and decays faster than 
$e^{-s}$ for $s\gs 1$. 
 Thus, ${{\zeta}}^*_{\alpha\beta}\sim A^{-3/2}$ for $\Omega\ll 1$ 
and $i\omega{{\zeta}}^*_{\alpha\beta}\sim 
1-(1-i)\sqrt{2/\Omega}$ 
for $\Omega \gg 1$.

The time-correlation  
function  ${{\hat R}}_{\alpha\beta}({\hat{\bi q}},t)
(\propto K_c(2t/\tau_D))$ 
decays  sharply  as a function of $t/\tau_D$, 
while  $R^*_{\alpha\beta}({\hat{\bi q}},\omega)$ has a  peak 
at $\omega=0$ with a  height $\propto  A^{-3/2}$ 
and a  width  $\propto A$. 
The  singularity of  $R^*_{\alpha\beta}({\hat{\bi q}},\omega)$ 
is described  by  the following function of $\omega$ and $A$,  
\be 
F^{\rm R}(\omega, A)= A^{-3/2} {\rm Re}[F_c(i\Omega)]  
= 16\Big[\Big(\sqrt{4A^2+\omega^2\tau_0^2}+2A\Big)^{1/2}-
2A^{1/2}\Big]/\omega^2\tau_0^2. 
\en  
which is equal to $A^{-3/2}$ at $\omega=0$  
and $A^{-3/2}/2$  at  $\omega\cong 4/\tau_D$. 
It  decays as $16/(\omega\tau_0)^{3/2}$ for $\Omega\gg 1$, 
which is independent of $A$. 
 In Fig.3a,  $F^{\rm R}(\omega, A)$  is displayed, which is 
much enlarged for $A\sim 10^{-3}$ and $\omega\ls 4/\tau_D$. 
Hereafter,  ${\rm Re}[\cdots]$ and ${\rm Im}[\cdots]$ 
denote taking the real part and the imaginary part of 
complex numbers, respectively.

 \begin{figure}[t]
\includegraphics[scale=0.92]{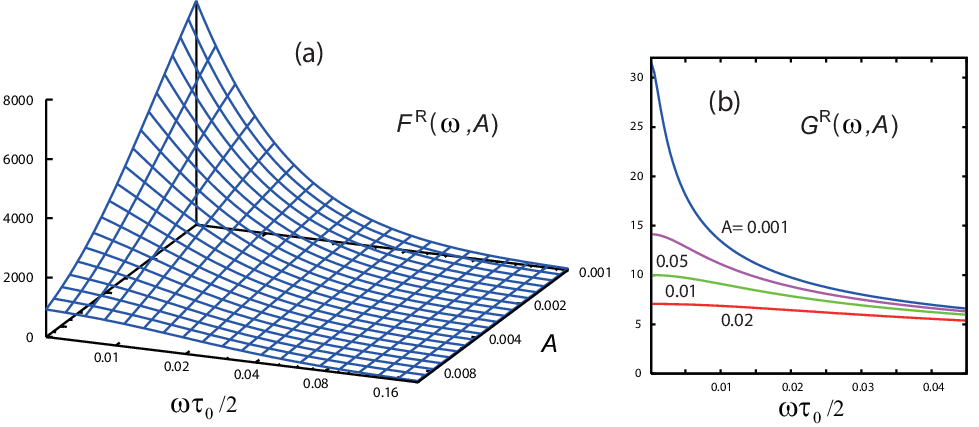}
\caption{\protect  
(a)  $F^{\rm R}(\omega, A)$ in Eq.(70) 
as a function of $\omega\tau_0/2$ and $ A$, 
which represents the near-critical strong growth of 
the central peak and the sound adsorption 
(see Eqs.(85) and (89)). 
(b)  $G^{\rm R}(\omega, A)$ in Eq.(86) 
vs $\omega\tau_0/2$ for three values of  $ A$, 
which gives the near-critical mild growth 
of the longitudinal sound velocity $v_{11}$ along $[001]$ 
in Eq.(84).  }
\end{figure}

The   real part of  $R^*_{\alpha\beta}({\hat{\bi q}},\omega)$ 
in Eq.(66) is the Fourier transform of 
${\hat R}_{\alpha\beta}({\hat{\bi q}},t)$ divided by 2  and is 
 proportional to $ F^{\rm R}(\omega, A))$ in Fig.3a. Thus, 
 it exhibits  the characteristic feature of    the   central peak 
observed  in  cubic BaTiO$_3$ \cite{Shira1,Ko,Ko3}. 
In   the light scattering experiment by   
Ko {\it el al.} \cite{Ko}, the  central peak 
 extended  over a broad  frequency range 
and its  {\it full width at half maximum } 
  was  given by  $ 
1037(T/T_0-1)$~GHz$=6.5(T/T_0-1)/$ps.
Their  finding    enables us 
to  determine $\tau_0$ in $\tau_D$. That is,  
since    ${\rm Re}[F_c(\Omega)]$  decays to 
$ 0.5$ at $\Omega=\omega\tau_D/2\cong 2$ in Fig.2b,   
we can equate their  width   to   $4/\tau_D$ to obtain     
\be 
\tau_D\cong  0.62/(T/T_0-1)~({\rm ps})\cong 
 1.1\times 10^{-2}A^{-1} ~({\rm ps}),  
\en
where we use $a_1$ in Eq.(6a). Thus,  $\tau_0\cong 
1.1\times 10^{-2}~$ps. However, if we use   $a_1$ in Eq.(6b), 
we have 
$\tau_0\cong  2.2\times 10^{-2} ~{\rm ps}$.   

We make remarks on other experiments 
on cubic BaTiO$_3$. (i) 
In their neutron scattering experiment, 
Yamada {\it et al.} \cite{Shira1} 
observed central peaks 
in the range $|\omega|\ls 1/$ps $(=0.6~$meV) 
at $T=410$~K, 
which is consistent with the result by Ko {\it et al.} \cite{Ko} 
for $T/T_c -1\sim 0.1$. (ii)  
In their hyper-Raman scattering experiment, 
  Vogt {\it et al.} \cite{Vogt}  obtained  $\tau_D=0.95$~ps 
at $T=446$~K, which is about one-third  of the value 
indicated by Eq.(71), however.

\vspace{2mm}
\noindent{\bf 4.3. Time-correlation 
of acoustic displacement and  elastic moduli}\\ 
Next, we consider   the dynamics of the 
acoustic displacement ${\hat{w}}_\alpha({\bi q},t)$ 
in Eq.(31). The   time-correlation 
of ${\hat{w}}_\alpha({\bi q},t)$  and its FL transform are 
written as  
\be
{\hat S}_{\alpha\beta}({\bi q},t)=
{ \av{{\hat{ w}}_\alpha ({\bi q},t)
{\hat{w}}_\beta({\bi q},0)^*}}/{V},~~~
S^*_{\alpha\beta}({\bi q},\omega)=\int_0^\infty dt e^{-i\omega t} 
{\hat S}_{\alpha\beta}({\bi q},t),
\en  
where ${\hat S}_{\alpha\beta}({\bi q},0)= k_BT\rho 
Z^{\alpha\beta}/q^2$ and  $S^*_{\alpha\beta}({\bi q},\omega)
\to  k_BT\rho Z^{\alpha\beta}/i\omega q^2$ as $\omega\to \infty$.  
In this paper, the total displacement time-correlation, written as  
${\hat U}_{\alpha\beta}({\bi q},t)$,  is approximated 
as the sum of the ES part ${\hat R}_{\alpha\beta}({\bi q},t)$ 
and the acoustic part ${\hat S}_{\alpha\beta}({\bi q},t)$ as  
\bea 
&&
\hspace{-10mm}
{\hat U}_{\alpha\beta}({\bi q},t)=
{ \av{{\hat{ u}}_\alpha ({\bi q},t)
{\hat{u}}_\beta({\bi q},0)^*}}/{V}
={\hat S}_{\alpha\beta}({\bi q},t)+ 
{\hat R}_{\alpha\beta}({\bi q},t), \\
&&\hspace{-10mm}
U^*_{\alpha\beta}({\bi q},\omega)=
\int_0^\infty dt e^{-i\omega t}
{\hat U}_{\alpha\beta}({\bi q},t) 
=S^*_{\alpha\beta}({\bi q},\omega)+ 
R^*_{\alpha\beta}({\bi q},\omega). 
\ena 
Here,   we neglect the cross time-correlation 
${ \av{{\hat{ m}}_\alpha ({\bi q},t)
{\hat{w}}_\beta({\bi q},0)^*}}
= { \av{{\hat{ w}}_\alpha ({\bi q},t)
{\hat{m}}_\beta({\bi q},0)^*}}$. This is justified if   
the time-evolution of $\bi p$ is not affected   
by $\bi u$ via the ES coupling. It  is indeed assumed 
 in   Eqs.(53) and (57).

In the linear response scheme \cite{Mori,Martin,Kawa,Gas1}, 
 $S^*_{\alpha\beta}({\bi q},\omega)$ 
 obeys the $3\times 3$ matrix equation,   
\be
{\sum}_\gamma 
 \Big[-{\rho\omega^2}\delta_{\alpha\gamma}
+q^2Z_{\alpha\gamma}^* ({\hat{\bi q}},\omega)\Big]
S^*_{\gamma\beta}({\bi q},\omega)=i\omega k_BT 
\rho Z^{\alpha\beta}/q^2, 
\en 
where   $\{ Z^*_{\alpha\gamma}({\hat{\bi q}},\omega) \}$ is 
the frequency-dependent Christoffel matrix. 
 written as 
\bea 
&&\hspace{-1cm} Z_{\alpha\beta}^* ({\hat{\bi q}},\omega)=
Z_{\alpha\beta}-\Psi_{\alpha\beta}
+i\omega \zeta^*_{\alpha\beta}(\omega)=
 Z_{\alpha\beta}+\int_0^\infty  dte^{-i\omega t} 
\frac{d}{dt } {\hat\Psi}_{\alpha\beta}({\hat{\bi q}}, t)\nonumber\\
&&\hspace{1cm}
=Z_{\alpha\beta}-\Psi_{\alpha\beta} G_c (i\Omega).
\ena 
Here,  $\zeta^*_{\alpha\beta}(\omega)$ is given 
in Eq.(64). In the second line we define  the   function 
  $G_c (i\Omega)$  by       
\bea 
&& \hspace{-15mm}G_c(i\Omega)=  1-{i\Omega }F_c(i\Omega)/4 
= 2(\sqrt{1+i\Omega}-1)/i\Omega\nonumber\\
&&\hspace{-2mm}= 
(2\sqrt{1+\Omega^2}-2)^{1/2}/\Omega -{i}\Big[
(2{\sqrt{1+\Omega^2}}+2)^{1/2}-{2}\Big]/\Omega,  
\ena  
where  the real and imaginary parts are written separately 
 as displayed  in   Fig.2c. 
Here, $G_c(i\Omega)\cong 1-i\Omega/4-5 \Omega^2/16$ 
for $\Omega\ll 1$ and  
  $G_c(i\Omega)\cong 2/\sqrt{i\Omega}=
(1-i) \sqrt{2/\Omega}$ 
for $\Omega\gg 1$. We can see 
 ${\rm Im}(G_c)<0$ leading  to positive entropy production.  
The sound dispersion relations with damping are 
given by vanishing of the determinant of the matrix 
$-{\rho\omega^2}\delta_{\alpha\beta}
+q^2Z_{\alpha\beta}^* $.

As a generalization of  Eq.(48), 
we use     
 Eqs.(26), (27), (40), and (76) 
to obtain   the renormalized   frequency-dependent elastic moduli,   
\bea 
&&\hspace{-10mm} 
C_{ij}^*(\omega) 
= c_{ij}-  M_{ij}I_eA^{-1/2}
G_c(i\Omega) = C_{ij}+
 M_{ij}I_eA^{-1/2}{i\Omega }F_c(i\Omega)/4 .  
\ena 
Writing   $C_{ij}^*(\omega)$ as $C_{ij}^* $, 
we find  the frequency-dependent Christoffel matrix,    
\be
Z_{\alpha\beta}^* ({\hat{\bi q}},\omega)
=(C^*_{44}+ (C^*_{11}-C^*_{12}-2C^*_{44})
{\hat q}_\alpha^2 \Big) 
 \delta_{\alpha\beta}+  (C^*_{12}+C^*_{44})
{\hat q}_\alpha{\hat q}_\beta, 
\en  
where $c_{ij}$ in $Z_{ij}$ in Eq.(26) are replaced by 
 $C_{ij}^* $. 
 From Eq.(78)  $C_{ij}^*(\omega) $ behave as   
\bea 
&&\hspace{2cm} C_{ij}^*(\omega) 
\cong  c_{ij}-  M_{ij}I_eA^{-1/2}(1-i \omega \tau_D/8) ~~~
(\omega\tau_D \ll 1) \nonumber \hspace{3.9cm}(80a) \\
&&\hspace{33mm} 
\cong c_{ij} -2 M_{ij}I_e (1-i)(\omega\tau_0)^{-1/2}~~~~~ 
(\omega\tau_D \gg 1), \nonumber \hspace{3.8cm}(80b)
\ena 
where  $C_{ij}^*(\omega) $ are independent of $A$ 
for $\omega\tau_D\gg 1$.   
\setcounter{equation}{80}
In particular, if ${\hat{\bi q}}$ is along $[001]$, 
we have 
${\hat q}_\alpha^2 = {\delta}_{\alpha z}$ and 
${\hat q}_\alpha{\hat q}_\gamma
={\delta}_{\alpha z}{\delta }_{\gamma z}$, so 
Eq.(75) is readily solved to give  
 \be 
S_{\alpha\beta}({\bi q},\omega)= k_BT 
\frac{i\omega\rho}{q^2}
\Big[\frac{\delta_{\alpha z}\delta_{\beta z}}{
c_{11}(-\rho\omega^2+ q^2C^*_{11})}+
 \frac{\delta_{\alpha\beta}- \delta_{\alpha z}\delta_{\beta z}
}{c_{44}(-\rho\omega^2+ q^2C^*_{44})}\Big]~~~~ 
({\hat{\bi q}} \parallel [001]). 
\en 
We will solve Eq.(75)  for general ${\hat{\bi q}}$ in   Appendix D.
.

From the cubic symmetry we find  the correlation-function expressions 
of $C_{ij}^*(\omega)$ as    
\bea 
&& \hspace{-16mm}C_{11}^*(\omega)=c_{11}+ 
{\cal G}_{zz}^{zz}(\omega), ~~~
C_{44}^*(\omega)=c_{44}+ 
{\cal G}_{zx}^{zx}(\omega), \nonumber\\
&&C_{12}^*(\omega)=c_{12}+\frac{1}{6}{\sum}_{\alpha\neq \beta} 
{\cal G}_{\alpha\alpha}^{\beta\beta}(\omega).
\ena 
Here,  we define the following correlation functions, 
\be 
{\cal G}_{\alpha\gamma}^{\beta\nu}(\omega)= 
\frac{1}{k_BT}
\int_0^\infty  \hspace{-1mm} dte^{-i\omega t} 
\frac{d}{dt } \int d{\bi r}
\av{\delta \Pi^{\rm s}_{\alpha\gamma}({\bi r},t) 
 \delta\Pi^{\rm s}_{\beta\nu}({\bi 0},0)} ,
\en 
which will be calculated in Appendix C in the mean-field theory. 
We note that    $\int d{\bi r}
\av{\delta \Pi^{\rm s}_{\alpha\gamma}({\bi r},t) 
 \delta\Pi^{\rm s}_{\beta\nu}({\bi 0},0)}/k_BT$ 
are the response functions in Kubo's   theory 
\cite{Kubo}.

\vspace{2mm}
\noindent{\bf 4.4. Acoustic anomaly due to 
electrostrictive coupling}\\ 
As can be known from Eq.(81), 
the displacement of  the longitudinal sound along $[001]$  
 is expressed as  const.$\exp[i\omega t - iq^*z]$ 
with  $ q^*= \omega (\rho/C_{11}^*)^{1/2}$ 
in the complex number representation. For this acoustic  mode,    
the  velocity $v_{11}= {\rm Re}(\omega/q^*)$ 
and the  absorption coefficient $\alpha^{\rm L}_{11}
= -{\rm Im}(q^*)$ 
are expressed in terms of $G_c(i\Omega)$ in Eq.(77) as 
\bea 
&& \hspace{-8mm} \frac{v_{11}}{v_{11}^0}=
1  - 
\frac{\Delta C_{11}}{2c_{11}} {\rm{Re}}[G_c(i\Omega)]
=1  - 
2.2\times 10^{-3}G^{\rm R}(\omega,A),
\\
&&\hspace{-10mm} \frac{v_{11}^0}{\omega}\alpha_{11}^{\rm L}= 
-\frac{\Delta C_{11}}{2c_{11}}
{\rm Im}[G_c(i\Omega)]
=2.7\times 10^{-4} \omega\tau_0  F^{\rm R}(\omega,A).
\ena 
Here,  $v_{11}^0=\sqrt{c_{11}/\rho}$ is the 
unrenormalized sound velocity and  
the experimental $\Delta C_{11}/c_{11}$ in the limit $\omega\to 0$ 
is given in Eq.(49).  
In Eq.(84)   we define the function, 
\be 
G^{\rm R}(\omega,A)= A^{-1/2}{\rm Re}[G_c(i\Omega)]= 
{2}\Big( \sqrt{4A^2+ \omega^2\tau_0^2} -2A\Big)^{1/2}
/\omega\tau_0,
\en 
which is plotted in Fig.3b. The  growth of $G^{\rm R}(\omega,A)$ 
at small $A$  is  much milder 
than that of $F^{\rm R}(\omega,A)$ in Fig.3a. 
Thus, $1-{v_{11}}/{v_{11}^0}\propto A^{-1/2}$ 
and  $\alpha^{\rm L}_{11}\propto \omega^2A^{-3/2}$ for $\Omega\ll 1$. 
However,  for  $\Omega\gg 1$, they are independent of $A$ as 
\be 
1-{v_{11}}/{v_{11}^0}\cong 
({v_{11}^0}/{\omega} ) \alpha^{\rm L}_{11}
\cong 0.43\times 10^{-2}({\omega\tau_0})^{-1/2}.
\en  
The above crossover occurs at very small $\omega$ 
for $A\ll 1$ (see Fig.3b), 

We comment on previous  theories on  
 the critical sound absorption, 
where the stress  tensor 
contains slowly relaxing terms quadratic   in  the 
order parameter.  
(i) Levanyuk  \cite{Leva2} assumed  the ES  
coupling  of the isotropic form 
 $\propto |{\bi p}|^2 \nabla \cdot{\bi u}$ 
(in the case of  $g_{11}-g_{12}= g_{44}=0$ in our Eq.(12)).
 He calculated the four-body time-correlation (FBTC) 
of $\bi p$. Indeed,  in  his Eqs.(19) and (20), 
he calculated  $G_c(i\Omega)$   in   our Eq.(77).  
(ii)   Minaeva {\it et al.} \cite{Leva1,Mina} 
assumed the ES coupling 
for uniaxial  TGS. They 
  calcualated FBTC  of the order parameter $p_y$ 
  using Krivoglaz's theory \cite{Kri2} 
to find  a critical growth of  the absorption coefficient 
($\propto A^{-1})$ in the low frequency limit. 
(iii) Dvo$\breve{\rm{r}}$$\acute{\rm a}$k  
\cite{Dvorak} assumed the general form of the 
ES  coupling for BaTiO$_3$ and  calculated FBTC,
 but he assumed its exponential relaxation 
not using $K_c(s)$ in our Eq.(68). 
(iv)  The sound  attenuation was studied 
 for ferromagnets in the asymptotic critical 
region by Kawasaki \cite{Kawasaki} 
and by  Laramore and Kadanoff \cite{Kada}. They 
calculated  FBTC  of the 
magnetization in the low frequency limit  in the mode-coupling 
theory, but they neglected the anisotropy 
in the crystal elasticity. (v) Analogously to 
$\zeta_{\alpha\beta}^*({\hat{\bi q}},\omega)$ in Eqs.(64) and (65), 
the bulk viscosity in near-critical fluids is   
  strongly divergent and  frequency-dependent, 
which is  given by  the FL transform 
of FBTC  of the order parameter (the mass density 
or the composition) \cite{Onukib,Kawa,Onuki1997}.

\begin{figure}[t]
\includegraphics[scale=0.9]{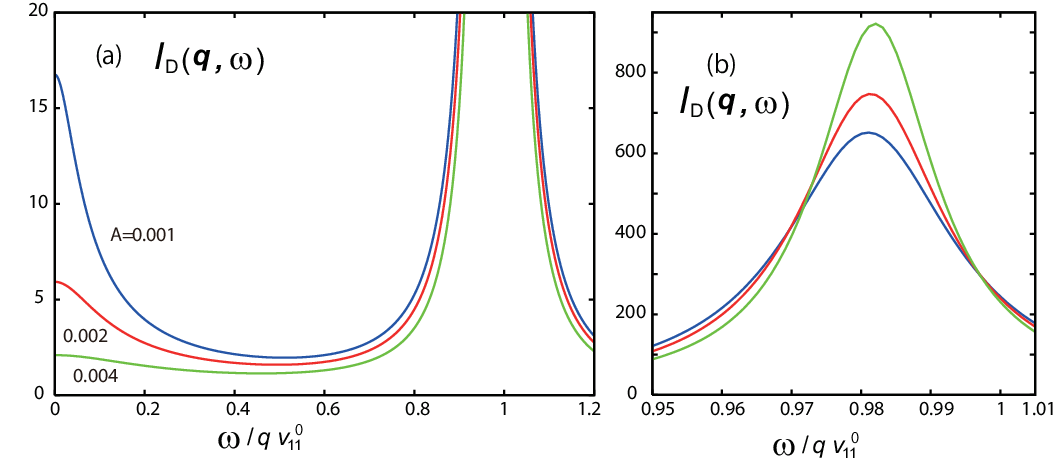}
\caption{\protect  
Normalized dynamic structure factor 
$(c_{11}/k_BT\tau_0)I_D(q,\omega)$ in Eq.(89) 
 vs scaled  frequency $\omega/qv_{11}^0$ along $[001]$ 
for $q=0.08~{\rm \AA}$ and $v_{11}^0=5.87\times 10^5$~cm$/s$. 
(a) The central peak is shown for $A=0.001$, $0.002$, 
and $0.004$ in the range $0<\omega/qv_{11}^0<1.2$, 
where $qv_{11}^0\tau_D= 0.052/A\gg 1$ for these $A$.  
 (b) The acoustic peak of $(c_{11}/k_BT\tau_0)I_D(q,\omega)$ 
  is shown   in the range $0.95<\omega/qv_{11}^0<1.01$.
 }
\end{figure}

\vspace{2mm}
\noindent{\bf 4.5. Density   time-correlation 
and dynamic structure factor}\\ 
We  consider   the FL transform of 
 the density time-correlation in  the limit $q\to 0$:  
\bea 
&&\hspace{-16mm}I^*_D({\bi q},\omega )=
\int_0^\infty dt e^{-i\omega t}
 {{\av{{\hat \rho} ({\bi q},t) 
{\hat \rho} ({\bi q},0)^*}}}/{{V}\rho^2}
= {\sum}_{\alpha,\beta}q_\alpha q_\beta U^*_{\alpha\beta}
({\bi q},\omega )
\nonumber\\
&& 
= {\sum}_{\alpha,\beta}q_\alpha q_\beta 
S^*_{\alpha\beta}({\hat{\bi q}},\omega)+ 
\frac{k_BT I_e\tau_0}{8(c_{12}+c_{44})^2 
} D({\hat{\bi q}})A^{-3/2}F_c(i\Omega),
\ena 
which tends to $ I_D({\bi q})/i\omega$ as $\omega \to \infty$ 
from Eq.(41). The acoustic part 
${\sum}_{\alpha,\beta}q_\alpha q_\beta 
S^*_{\alpha\beta}$ for general $\hat{\bi q}$ will be calculated 
 in  Appendix D. In the ES part,     $D({\hat{\bi q}})$ 
 is given  in Eq.(43) and  Fig.1. 
The real part $I_D({\bi q},\omega)={\rm Re}[I^*_D({\bi q},\omega )]$ 
is called the dynamic structure factor.  
It exhibits the Brillouin and central peaks, 
whose  $\omega$-integrals in the range $\omega>0$ 
 are the two terms in $I_D({\bi q})$ 
in Eq.(42) multiplied by $\pi$. 
 Here, using  Eq.(81), 
we calculate   $I_D({\bi q},\omega )$   
for $\bi q$ along $[001]$:   
\be 
\frac{c_{11}}{k_BT\tau_0}I_D({\bi q},\omega)=
  {\rm Re}\Big(
\frac{-\rho i\omega/\tau_0}{\rho \omega^2-q^2C_{11}^*}\Big)
+\frac{c_{11} I_eD[001]}{8(c_{12}+c_{44})^2 
} F^{\rm R}(\omega,A)  
  ~~~~({\bi q}\parallel[001]),
\en 
where $D[001]=2.8$ in Fig.1, $F^{\rm R}(\omega,A)$ is 
given in Eq.(70), and the coefficient in front of 
$F^{\rm R}(\omega,A)$ is 
estimated to be  $5.4\times 10^{-3} $ from Table 1 and Eq.(71). 
The  height and width of the acoustic part  
are  $1/(2v_{11}^0\alpha_{11}^{\rm L}\tau_0)$ 
and $v_{11}^0\alpha_{11}^{\rm L}$, respectively, 
in terms of  $\alpha_{11}^{\rm L}$ in Eq.(85),   
while those of the ES part  are $5.4\times 10^{-3} A^{-3/2}$ 
and $4/\tau_D$, respectively. 

In Fig.4, we plot   
$I_D({\bi q},\omega)$ along $[001]$ 
vs $\omega/qv_{11}^0$, where  $q=0.08/{\rm \AA}$, 
  $v_{11}^0= 5.87\times 10^5$ cm$/$s, and 
 $qv_{11}^0= 4.7/$ps$(= 750$~GHz$=2.9$~meV). 
This $q$ value  was used in 
 the neutron scattering 
experiment by Yamada {\it et al} \cite{Shira1}. 
For the three curves in Fig.4,   we set  $A=0.001$, $0.002$, 
and $0.004$ using   $a_1=1.8\times 10^{-2}$ in   Eq.(6a), 
for which $qv_{11}^0\tau_D= 52,~ 26,$ and $13$, respectively. 
Thus, in Fig.4,      the acoustic frequency  $qv_{11}$ exceeds  the central 
peak width $4/\tau_D$. We also notice that 
 $v_{11}$ and $\alpha_{11}^{\rm L} $  are  insensitive to $A$ 
from  Eq.(87). We can also see that the peak height 
of the central peal is much lower  than that of the acoustic peak 
for $T>T_c$. From Eqs.(85) and (89) the ratio of these heights is 
calculated as  
\bea 
&&\frac{{\rm  central~ peak~height}}{{\rm 
acoustic~ peak~height}}=4.8\times 10^{-6}
\Big[ \Big(\sqrt{4+(qv_{11}^0\tau_0)^2/A^2}+2\Big)^{1/2}-2\Big]/A,
\ena 
where grows as $A\to 0$ but the coefficient is very small. 
For $A\ll qv_{11}^0\tau_0$, 
 the above ratio becomes 
  $4.8\times 10^{-6}\sqrt{qv_{11}^0\tau_0} A^{-3/2}$, 
which is the case in  Fig.4.  

We make remarks. (i) In the 
neutron scattering experiment by Yamada {\it et al.}
the central peak decreased with further increasing $q$ 
to $0.12/{\rm \AA}$ and $0.16/{\rm \AA}$ in their Fig.7. This large-$q$ 
behavior is beyond the scope of this paper since 
the limit $q\to 0$ is taken in our theory. 
(ii) Along $[001]$,  Ko {\it et al.} \cite{Ko} 
set $q\cong 4\times 10^{-3}/{\rm \AA}$ 
in measuring  the Brillouin shift (in their Fig.1) 
and  $q\cong 4\times 10^{-2}/{\rm \AA}$ 
in measuring  the central peak in their Fig.4, 
where the acoustic frequency is 
at most comparable to  the central peak width. 

\vspace{2mm}
\noindent{\bf 5. Summary and remarks}\\ 
We have presented a Ginzburg-Landau 
theory of the central peak and the acoustic anomaly 
in  paraelectric   BatTiO$_3$, which originate  from 
the electrostrictive (ES) coupling between 
the polarization $\bi p$ and the displacement $\bi u$. 
We have used the mean-field theory, 
which is justified by the Ginzburg criterion in Sec.3.1.  
Our main results are summarized as follows.

In Sec.2, we have presented the free energy  functional 
$\cal F$, where the  ES  stress tensor $\Pi^{\rm s}_{\alpha\beta}$ 
in Eqs.(11) and (12) is a key quantity. 
The total stress tensor has been given in Eq.(16).

In Sec,3, the static correlations of $\bi p$ and $\bi u$ 
have been calculated. 
In Sec.3.2, those of  the 
longitudinal and transverse Fourier components of $\bi p$ 
have been given in Eq.(28). In Sec.3.3, 
$\bi u$ has been divided into the ES  part $\bi m$ 
and the acoustic part ${\bi w}$ in Eq.(31) with 
 their  correlations 
 in Eq.(37). The ES  correlation 
$R_{\alpha\beta}({\bi q})$ grows as $A^{-1/2}$ 
in the limit $q\to 0$ and is related  
 to the ES coefficients $g_{ij}$ 
 in Eqs.(38)-(40), In Sec.3.4, we have also examined  the density 
correlation $I_D({\bi q})$  in Eq.(42), 
whose ES  part is proportional to the 
anisotropy  factor $D({\hat{\bi q}})$ in Fig.1. 
In Sec.3.5, we have calculated the renormalized static 
elastic moduli $C_{ij}$ 
 in Eq.(48). We have then  determined the coefficient $C$ 
in the gradient free energy  in Eq.(50) using  
the data by Kashida {\it et al.} \cite{Kashida}. 
In Sec.3.6, we have obtained  the free energy 
contribution $\Delta{\cal F}^{\rm po}_{\rm inh}$ 
  from the inhomogeneous displacement fluctuations in Eq.(51). 

In Sec.4-1, starting with 
 the dynamic equations for $\bi p$, we have 
 calculated the  time-correlation function of $\bi p$  
in Eq.(57). The  effective dielectric constant 
 for films has been given in Eqs.(59) and (60).  
   In Sec.4-2, we have examined the time-correlations  
of the ES   displacement, ${\hat R}_{\alpha\beta}({\bi q},t)$ 
and $R^*_{\alpha\beta}({\bi q},\omega)$, 
 in Eqs.(61) and (62) in terms of  the 
stress time correlations  in Eqs.(61) and (62). 
These  functions 
involve  the Debye relaxation time $\tau_D=\tau_0/A$ 
and depend on $t$ and $\omega$ simply as in Eqs.(62) and (63). 
The ES correlation $R^*_{\alpha\beta}({\bi q},\omega)$ 
has the characteristic feature of the central peak.  
as is evident  in Fig.3a. 
We have then determined 
$\tau_D$ in Eq.(68) using the data by Ko {\it et al.} \cite{Ko1}. 
 In Sec.4-3, we have calculated  the time-correlations  
of the acoustic  displacement, ${\hat S}_{\alpha\beta}({\bi q},t)$ 
and $S^*_{\alpha\beta}({\bi q},\omega)$, 
 in Eq.(69). We have then obtained 
the frequency-dependent elastic moduli 
$C^*_{ij}(\omega)$ in Eqs.(78) and (80), where 
the critical part of   $C^*_{44}(\omega)$ is very small. 
   In Sec.4-4, we have 
examined the  acoustic anomaly of the longitudinal sound 
along $[001]$ in  Eqs.(84)-(87). 
 In Sec.4-5, the density time-correlation 
$I_D({\bi q},\omega)$ in Eq.(89) 
has been displayed in Fig.4 in accord with the 
 scattering experiment by Yamada {\it et al.} 
\cite{Shira1}.

We remark on future work. 
(i) We will shortly present a theory on 
the central peak in tetragonal  uniaxial 
 KDP, where the piezoelectric stress 
 is given by  $g_{36} p_z (\delta_{\alpha x}\delta_{\beta y}
+ \delta_{\alpha y}\delta_{\beta x})$ 
with $g_{36}$ being a constant. 
(ii) We should   also present a theory 
below $T_c$. (iii)  Domain dynamics 
in various ferroelectrics 
is of great interest \cite{Chen2,Nam}, 
where  an  electric field can be applied.   
(iv) In systems with large dielectric 
constants, the statics and dynamics  generally  
 depend on whether the  surface charge $Q_0$ is fixed 
or the potential difference $\Phi_a$ is fixed 
 (see below Eq.(58)) \cite{Sprik1,Sprik,Onuki2025}.  
 That is,  at fixed $\Phi_a$, 
the space-time polarization correlations 
acquire nonlocal parts inversely 
proportional to the volume $V$, 
whose sample integral 
determines the  overall  dielectric 
response.

\vspace{2mm} 
  \noindent 
{\bf Data availability}\\
 The data that supports the findings of
this study are available within the article.

\vspace{2mm}
\noindent{\bf Appendix A: Global mechanical equilibrium 
in  unclamped samples  }\\
\setcounter{equation}{0}
\renewcommand{\theequation}{A\arabic{equation}}
In   unclamped samples, homogeneous strains can be 
 induced by overall  polarization. Thus, we write the sample averages 
 of   $\epsilon_{\alpha\beta}$, $\sigma_{\alpha\beta}$, 
$\Pi_{\alpha\beta}^{\rm s}$,  and $\epsilon_{\alpha\beta}^{\rm s}$ 
as $\overline{{\bar \ep}_{\alpha\beta}}$, 
 $\ov{{\sigma}_{\alpha\beta}}$, $\ov{{\Pi}_{\alpha\beta}^{\rm s}}$, 
 and $\ov{{\ep}_{\alpha\beta}^{\rm s}}$, respectively, with overlines.  
They are the Fourier components with ${\bi q}={\bi 0}$ 
divided by $V$ (see Eq.(23)). 
For example, we can prepare  a stress-free  
paraelectric state with ${\bi u}={\bi p}={\bi 0}$ 
and  apply a small electric field. Then, these 
sample averages become nonvanishing. 
Their   contribution  to 
$\int d{\bi r}(f_{\rm el}  + f_{\rm int})$ is written as  
\be 
{\Delta{\cal F}_{\rm hom}^{\rm el}} =
V{\sum}_{\alpha,\beta}\Big(
~\frac{1}{2}\ov{{\sigma}_{\alpha\beta}}
\cdot \ov{{\epsilon}_{\alpha\beta}}
 -\ov{\Pi_{\alpha\beta}^{\rm s}}\cdot\ov{{\epsilon}_{\alpha\beta}}
\Big), 
\en   
which is minimized with respect to $\overline{{\bar \ep}_{\alpha\beta}}$ 
in the  mechanical equilibrium at given 
 $\ov{{\Pi}_{\alpha\beta}^{\rm s}}$. Then, 
\be 
\ov{{{\ep}}_{\alpha\beta}}= \ov{{{\ep}}_{\alpha\beta}^{\rm s}}, 
~~~~
\ov{{{\sigma}}_{\alpha\beta}}= \ov{{{\Pi}}_{\alpha\beta}^{\rm s}},
 ~~~~
{\Delta{\cal F}_{\rm hom}^{\rm el}} =-\frac{1}{2}V
 {\sum}_{\alpha,\beta}
\ov{{{\Pi}}_{\alpha\beta}^{\rm s}}\cdot \ov{{{\ep}}_{\alpha\beta}^{\rm s}},   
\en  
where the first two relations are equivalent. Using Eq.(A2) 
  the ES  coefficients $Q_{ij}$ in Eq.(13) 
can be determined  experimentally \cite{Cross,Chen1,Uchino}. 
 From Eqs.(12)-(14) we obtain 
\be
{\Delta{\cal F}_{\rm hom}^{\rm el}}/V =
 -\beta_1{\sum}_\alpha 
\Big(\ov{p_\alpha^2}-\frac{1}{3}\ov{|{\bi p}|^2}\Big)^2
  -\beta_2 (\ov{|{\bi p}|^2})^2- \beta_3
\Big[(\ov{p_xp_y})^2+(\ov{p_yp_z})^2+ (\ov{p_zp_x})^2\Big] .
\en 
The  coefficients $\beta_1$, $\beta_2$, and 
 $\beta_3$  are positive and are given by 
\be   
\beta_{1}={(g_{11}-g_{12})^2}/[{2(c_{11}-c_{12})]},~~
\beta_{2}={(g_{11}+2g_{12})^2}/[{6(c_{11}+2c_{12})}],~~
\beta_3= {g_{44}^2}/{2c_{44}} .
\en 
For BaTiO$_3$,    Table 1 gives  $\beta_1=0.030$, $\beta_2=0.0026$, 
and $\beta_3=0.0014$  in ${\rm  GPa}^{-1}$. 

In Eq.(A3) the overlines can be  removed for stress-free 
  single crystals with homogeneous $\bi p$. 
After  the homogeneous strains are removed 
in mechanical equilibrium in such samples,  
 $\alpha_{11}^0$ and $\alpha_{12}^0$ in Eq.(3) are changed to 
\be 
\alpha_{11}=\alpha_{11}^0
-2\beta_1/3- \beta_2, ~~\alpha_{12}=\alpha_{12}^0
+2\beta_1/3-2\beta_2-\beta_3, 
\en 
For stress-free  BaTiO$_3$,   
some authors \cite{Renoud,Bell,Cross}  
determined the  Landau free energy  for  
homogeneous $\bi p$ to explain  the observed successive 
first-order phase transitions, where   
 ${\Delta{\cal F}_{\rm hom}^{\rm el}}$ in Eq.(A3) 
 should be   included in the free energy. 
Li {\it et al} \cite{Cross}   found 
 $\alpha_{12}= 
7.974\times 10^8~{\rm C}^{-4}{\rm m}^6{\rm N}
= 0.010/{\rm GPa}=3.05/c_{11}>0$ 
and $\alpha_{11}=-0.26\alpha_{12}<0$, 
where $\alpha_{12}$  is one-third of $\beta_1$ in Eq.(A4).  
Then,  Eq.(A5) gives $\alpha_{11}^0= 0.025/$GPa$>0$  
and  $\alpha_{12}^0=- 0.006/$GPa$<0$.  
In BaTiO$_3$, the  phase transitions  
 delicately  depend  on 
the mechanical boundary condition.  

We  also note  that the ovelines  in Eq.(A3) 
cannot be removed for inhomogeneous $\bi p$,  
 particularly, in multi-domain states and in phase ordering. 
This aspect remains  unexplored.

\vspace{2mm}
\noindent{\bf Appendix B: Gradient free energy for cubic solids }\\
\setcounter{equation}{0}
\renewcommand{\theequation}{B\arabic{equation}}
In the literature \cite{Chen2,Nam,Cross1}, 
the gradient free energy ${\cal F}_g$ 
for BaTiO$_3$ consists of several terms, which are  
bilinear in $\nabla_\alpha p_\beta$ 
with four coefficients $G_{11}, G_{12}, G_{44}$, and $G_{44}'$. 
We express  ${\cal F}_g$ 
in terms of  the Fourier components 
${\hat{p}}_\alpha({\bi q})$ 
with ${\bi q}\neq {\bi 0}$ in Eq.(23)  as 
\be 
{\cal F}_g= 
 \frac{1}{2V}{\sum}_{{\bi q}\neq {\bi 0}}C\Big[
(1-{\gamma}_{14}-{\gamma}_{44}){\sum}_\alpha q_\alpha^2 
|{\hat{p}}_\alpha({\bi q})|^2+ {\gamma}_{14}
|{{\bi q}\cdot\hat{\bi p}}({\bi q})|^2
+{\gamma}_{44} q^2|{\hat{\bi p}}({\bi q})|^2\Big],    
\en 
where   $C$, ${\gamma}_{14}$, 
 and ${\gamma}_{44}$ are given by 
   $C= G_{11}-G_{44}$, $C\gamma_{14}= G_{12}+G_{44}-G_{44}'$, 
and $G\gamma_{44}= G_{44}+G_{44}'$. 
Here, we do not write  the surface contributions 
 $\propto \int dxdy(p_\alpha \nabla_z p_\beta)_{z=0,H}$, 
which are relevant  for thin films in the presence 
of the surface free energy \cite{Binder}. 
  
In simulations, Nambu and Sagara \cite{Nam} 
set $({\gamma}_{14},{\gamma}_{44})=(1.9,1)$ 
in two dimensions, while Hu and Chen \cite{Chen2}  set 
$({\gamma}_{14},{\gamma}_{44})=(0,1)$ in three dimensions. 
For the latter choice, we  obtain the isotropic form 
${\cal F}_g={\sum}_{{\bi q}\neq {\bi 0}}
Cq^2|{\hat{\bi p}}({\bi q})|^2/2V$ 
 adopted in   this paper.

\vspace{2mm}
\noindent{\bf Appendix C: Correlations of 
electrostrictive tensor $\Pi_{\alpha\beta}^{\rm s}$ at long wavelengths }\\
\setcounter{equation}{0}
\renewcommand{\theequation}{C\arabic{equation}}
In this appendix, we express  the  time-correlation of any 
space-time-dependent variables ${\cal A}({\bi r},t)$ 
and  $ {\cal B}({\bi r},t)$  in the long wavelength 
limit  as   
\be 
\av{{\cal A}:{\cal B}}(t) = \int\hspace{-1mm}  d{\bi r} 
({\cal A}({\bi r},t)- \av{{\cal A}})({\cal B}({\bi 0},0)- \av{{\cal B}}),
\en 
where  the integrand  is assumed to decay rapidly 
for large $r$.   Using Eq.(57) we   calculate 
 the four-body   time correlation 
of  $ m_{\alpha\gamma}= 
p_\alpha p_\beta$ and  $ m_{\beta\nu}= 
p_\beta p_\nu$ as  
\bea 
&&\hspace{-27mm} 
\av{m_{\alpha\gamma} :m_{\beta\nu}}(t )=((2\pi)^{-3}   
\int {d{\bi q}}~  [G_{\alpha\beta}({\bi q},t)G_{\gamma\nu}({\bi q},t)
+ G_{\alpha\nu}({\bi q},t)G_{\gamma\beta}({\bi q},t)]\nonumber\\
&& =  [
{7}(\delta_{\alpha\beta}
\delta_{\gamma\nu}+\delta_{\alpha\nu}\delta_{\beta\gamma})
+ {2}\delta_{\alpha\gamma}\delta_{\beta\nu}]J_e(t),
\ena 
where we use the  decoupling approximation and  
pick up the contributions  
proportional to  $\int d{\bi q}{\hat G}_\perp(q,t)^2$.
Using  $I_e$ in  Eq.(30) and 
$K_c(s)$   in Eq.(68)  we calculate $J_e(t)$  as    
\be 
J_e(t) =  (k_BT)^2 
\int \hspace{-1mm} d{\bi q}~G_\perp(q,t)^2/120\pi^3
= k_BT I_e A^{-1/2}K_c (2t/\tau_D)/60.
\en

It is convenient to introduce the following variables 
quadratic in $\bi p$:   
\be 
{\cal D}_\alpha= p_\alpha^2- |{\bi p}|^2/3, ~~~~
{\cal E}_{\alpha\beta}=p_{\alpha}p_{\beta}- p_\alpha^2\delta_{\alpha\beta}.
\en   
From Eq.(C2)  
$|{\bi p}|^2$, ${\cal D}_\alpha$, and ${\cal E}_{\alpha\beta}$ 
are {\it orthogonal} to one another as
\bea 
&&\hspace{-12mm} 
\av{{\cal D}_{\alpha}: 
{\cal E}_{\beta\nu}}(t)
=\av{ {\cal D}_{\alpha} :{|{\bi p}|^2}}(t) 
=\av{ {\cal E}_{\alpha\beta}: |{\bi p}|^2}(t) =0.
\ena 
Their self-correlations are  given by 
\bea 
&& \hspace{-2mm}\av{ {|{\bi p}|^2}: |{\bi p}|^2}(t) =60J_e(t),~~
\av{ {\cal D}_{\alpha}:  {\cal D}_{\beta}}(t) 
=14(\delta_{\alpha\beta}-1/{3})J_e(t),\nonumber\\
&&\hspace{-2mm} 
\av{{\cal E}_{\alpha\gamma}: {\cal E}_{\beta\nu}}(t) = 
7(\delta_{\alpha\beta}\delta_{\gamma\nu}+\delta_{\alpha\nu}
\delta_{\beta\gamma}-2
\delta_{\alpha\beta}\delta_{\gamma\nu}\delta_{\beta\nu})J_e(t).
\ena 

We then   express the  
stress tensor $\Pi_{\alpha\beta}^{\rm s}$ in Eq.(12) as 
\be 
\Pi_{\alpha\beta}^{\rm s}= 
(g_{11}-g_{12}){\cal D}_\alpha \delta_{\alpha\beta} 
+ (g_{11}+2g_{12})
{|{\bi p}|^2}\delta_{\alpha\beta}/3
+ g_{44}{\cal E}_{\alpha\beta}. 
\en 
Using Eqs.(C5) and (C6)   we find 
the time-correlations of the ES stress tensor, 
\bea 
&& \hspace{-11mm} 
 \av{ \Pi^{\rm s}_{\alpha\gamma}: 
 \Pi^{\rm s}_{\beta\nu}}(t) \hspace{-1mm}= 60J_e(t)  
{\cal T}_{\alpha\gamma}^{\beta\nu}
= k_BT I_e A^{-1/2}{\cal T}_{\alpha\gamma}^{\beta\nu}
K_c (2t/\tau_D),\\
&&\hspace{-1mm}
{\cal G}_{\alpha\gamma}^{\beta\nu}(\omega)= 
 I_e A^{-1/2} {\cal T}_{\alpha\gamma}^{\beta\nu} G_c(i\Omega),
\ena 
where we define 
${\cal G}_{\alpha\gamma}^{\beta\nu}(\omega)$ in 
Eq.(83),  $G_c(i\Omega)$ in Eq.(77), and 
${\cal T}_{\alpha\gamma}^{\beta\nu}$  as  
\bea 
&&\hspace{-40mm}
{\cal T}_{\alpha\gamma}^{\beta\nu}=
  \delta_{\alpha\gamma}\delta_{\beta\nu}(M_{11}+2M_{12})/3
+(\delta_{\alpha\beta}-1/3)\delta_{\alpha\gamma}\delta_{\beta\nu}
(M_{11}-M_{12})\nonumber\\
&&\hspace{-20mm}
+(\delta_{\alpha\beta}\delta_{\gamma\nu}+\delta_{\alpha\nu}
\delta_{\beta\gamma}-2
\delta_{\alpha\beta}\delta_{\gamma\nu}\delta_{\beta\nu})M_{44}.
\ena 
From Eq.(39) we have $\Psi_{\alpha\beta}({\bi q})= I_eA^{-1/2} 
{\sum}_{\gamma,\nu}{\hat{q}}_\gamma {\hat{q}}_\nu 
{\cal T}_{\alpha\gamma}^{\beta\nu}$. 
We can then confirm Eqs.(40), (65), (66), 
and (82)  from Eqs.(C8)-(C10). 

\vspace{2mm}
\noindent{\bf Appendix D: Fourier-Laplace  transform of 
acoustic time-correlation }\\
\setcounter{equation}{0}
\renewcommand{\theequation}{D\arabic{equation}}
To calculate $S^*_{\alpha\beta}({\bi q},\omega)$ 
for general $\hat{\bi q}$  from  Eqs.(75) and (79),  we  
 define    $a_\alpha^*$, $b^*$, and ${W^*_{\alpha\beta}}$ by 
\bea 
&& \hspace{-10mm}a_{\alpha}^*= -\rho\omega^2/q^2+ C_{44}^*+ 
(C_{11}^*-C_{12}^*-2C_{44}^*){\hat{q}}_\alpha^2,~~
b^*=(C_{12}^*+C_{44}^*){\sum}_\gamma {\hat{q}}_\gamma^2/a_\gamma^*,
\nonumber\\
&&\hspace{-6mm}{W}^*_{\alpha\beta}=  
 { \delta_{\alpha\beta}}/{a_\alpha^*}
-{(C^*_{12}+C^*_{44})
{\hat q}_\alpha{\hat q}_\beta}/[{a_\alpha^* a_\beta^*(1+b^*)}],
\ena 
Here,  the matrix $\{W^*_{\alpha\beta}\}$ is 
the inverse of  $\{-{\rho\omega^2}\delta_{\alpha\beta}/q^2
+Z_{\alpha\beta}^*\}$. It  is 
obtained from 
$Z^{\alpha\beta}$ in Eq.(32)  by replacements: $a_\alpha\to a_\alpha^*$ 
and $b\to b^*$. 
Then, Eq.(75) yields 
\be
S_{\alpha\beta}^* ({\bi q},\omega)= \frac{\rho k_BT i\omega}{ q^4}{\sum}_\gamma
{W^*_{\alpha\gamma}}{Z^{\gamma\beta}}= 
\frac{\rho k_BT i\omega}{a_\alpha^*a_\beta q^4}
\Big[\delta_{\alpha\beta} - {{\hat q}_\alpha {\hat q}_\beta}
\Xi_{\alpha\beta}({\hat{\bi q}})\Big], 
\en 
where $\Xi_{\alpha\beta}$ is given  by 
\be 
 \Xi_{\alpha\beta}({\hat{\bi q}})=\frac{C^*_{12}+C^*_{44}}{a_\beta^*(1+b^*)}
+\frac{C_{12}+C_{44}}{a_\alpha(1+b)}
- \frac{(C^*_{12}+C^*_{44})(C_{12}+C_{44})}{(1+b^*)(1+b)}
{\sum}_\gamma \frac{{\hat q}_\gamma^2}{a_\gamma^* a_\gamma} .
\en 
Using ${\sum}_\gamma {W^*_{\alpha\gamma}}{\hat q}_\gamma=
{\hat q}_\alpha /a_\alpha^* (1+b^*)$ 
 the acoustic part in Eq.(88)  is expressed   as 
\be 
{\sum}_{\alpha,\beta}q_\alpha q_\beta S_{\alpha\beta}^*
  ({\bi q},\omega)=\frac{\rho k_BT i\omega}{(1+b^*)(1+b)}
{\sum}_\gamma \frac{{\hat q}_\gamma^2}{a_\gamma^* a_\gamma}. 
\en 
 

\end{document}